\newif\ifpreview

\documentclass[fleqn,11pt,onecolumn]{wlscirep}
\usepackage[utf8]{inputenc}
\usepackage[T1]{fontenc}
\usepackage{graphicx}
\usepackage{setspace}
\usepackage[version=3]{mhchem}
\usepackage{lineno}
\usepackage{bm}
\usepackage{amsmath}
\usepackage{amsfonts}
\usepackage{mathrsfs}
\usepackage{color}
\usepackage{xurl}
\usepackage[abs]{overpic} 
\usepackage{subcaption}

\usepackage{booktabs}
\usepackage{multirow}
\usepackage[subrefformat=parens]{subcaption}

\ifpreview
\newcommand{\rem}[1]{\textcolor{lightgray}{#1}}

\newcommand{\note}[1]{\textcolor{red}{[#1]}}

\else
\newcommand{\rem}[1]{\textcolor{lightgray}{}}

\newcommand{\note}[1]{}
\fi

\newcommand{\ie}{\emph{i.e.}}
\newcommand{\eg}{\emph{e.g.}}
\newcommand{\todo}[1]{}
\renewcommand{\todo}[1]{{\color{red} TODO: {#1}}}

\title{Neural Structure Fields with Application to Crystal Structure Autoencoders}

\author[1, 2]{Naoya Chiba}
\author[3, 4]{Yuta Suzuki}
\author[1]{Tatsunori Taniai}
\author[1]{Ryo Igarashi}
\author[1]{Yoshitaka Ushiku}
\author[5,6]{Kotaro Saito}
\author[6,*]{Kanta Ono}  

\affil[1]{OMRON SINIC X Corporation, Nagase Hongo Building 3F, 5-24-5 Hongo, Bunkyo, Tokyo, Japan.}
\affil[2]{Graduate School of Information Sciences, Tohoku University, 6-6-01 Aramaki-Aza-Aoba, Aoba, Sendai, Miyagi, Japan.}
\affil[3]{TOYOTA Motor Corporation, 1200, Mishuku, Susono, Shizuoka, Japan.}
\affil[4]{The Graduate University for Advanced Studies (SOKENDAI), 1-1 Oho, Tsukuba, Japan.}
\affil[5]{Randeft Inc. TOKIWA Bridge 13F, 2-7-4, Otemachi, Chiyoda, Tokyo, Japan.}
\affil[6]{Department of Applied Physics, Graduate School of Engineering, Osaka University, 2-1 Yamadaoka, Suita, Osaka, Japan.}
\affil[*]{ono@ap.eng.osaka-u.ac.jp}

\begin{abstract}
Representing crystal structures of materials to facilitate determining them via neural networks is crucial for enabling machine-learning applications involving crystal structure estimation. Among these applications, the inverse design of materials can contribute to explore materials with desired properties without relying on luck or serendipity. We propose neural structure fields (NeSF) as an accurate and practical approach for representing crystal structures using neural networks. Inspired by the concepts of vector fields in physics and implicit neural representations in computer vision, the proposed NeSF considers a crystal structure as a continuous field rather than as a discrete set of atoms. Unlike existing grid-based discretized spatial representations, the NeSF overcomes the tradeoff between spatial resolution and computational complexity and can represent any crystal structure. We propose an autoencoder of crystal structures that can recover various crystal structures, such as those of perovskite structure materials and cuprate superconductors. Extensive quantitative results demonstrate the superior performance of the NeSF compared with the existing grid-based approach.
\end{abstract}

\begin{document}
\ifpreview
\maketitle
\else
\maketitle
\thispagestyle{empty}
\fi

\section{Introduction}
A fundamental paradigm in materials science considers structure--property relationships assuming that the material properties are tightly coupled with their crystal structures. Thus, for conventional approaches in materials science, theoretical and experimental analyses of the structure--property relationships of materials are conducted in the search for novel materials with superior properties\cite{degraefStructureMaterials2012, callisterMaterialsScienceEngeneering2010}. However, these conventional approaches  rely on labor-intensive human analysis  and even ``serendipity.''

To automate or assist material analysis and development, data-driven approaches have been actively studied in materials science, establishing the area of materials informatics (MI)\cite{lookmanInformationScienceMaterials2015, schmidtRecentAdvancesApplications2019, szymanskiAutonomousDesignSynthesis2021a, choudharyRecentAdvancesApplications2022}. Unlike conventional approaches based on the deduction of physical laws, MI aims to unveil materials knowledge (\eg, laws governing structure--property relationships) from datasets of collected materials via statistical and machine learning (ML) methods. In recent years, MI has been developed rapidly owing to technological advances in ML and the advent of large-scale materials databases\cite{agrawalPerspectiveMaterialsInformatics2016, choudharyRecentAdvancesApplications2022}. Thus, powerful neural-network-based ML methods are becoming key components in MI research\cite{choudharyRecentAdvancesApplications2022}. Applications of MI include the prediction of material properties from material characteristic data such as crystal structures\cite{choudharyRecentAdvancesApplications2022,suzukiSelfSupervised2022} and compositions\cite{jhaElemNetDeepLearning2018,goodallPredictingMaterialsProperties2020, parkDevelopingImprovedCrystal2020, wangCompositionallyRestrictedAttentionbased2021}, automated analyses of experimental data\cite{parkClassificationCrystalStructure2017,oviedoFastInterpretableClassification2019, szymanskiProbabilisticDeepLearning2021}, and natural language processing for knowledge retrieval from scientific literature\cite{tshitoyanUnsupervisedWordEmbeddings2019}.

Many MI studies have been focused on predicting properties of given materials\cite{xieCrystalGraphConvolutional2018, chengGeometricinformationenhancedCrystalGraph2021, choudharyAtomisticLineGraph2021}, such as the bandgap, Seebeck coefficient, and elastic modulus. 
When this type of task is viewed as the discovery of \emph{structure-to-property relationships} among materials, there is yet another important type of task, namely, the discovery of \emph{property-to-structure relationships}, which constitutes an inverse design problem\cite{nohInverseDesignSolidState2019, nohMachineenabledInverseDesign2020a, yaoInverseDesignNanoporous2021, longConstrainedCrystalsDeep2021, fungInverseDesignTwodimensional2021}. 
Despite the great potential utility of this inverse approach in developing materials, few studies have addressed this\cite{nohInverseDesignSolidState2019, nohMachineenabledInverseDesign2020a, yaoInverseDesignNanoporous2021, longConstrainedCrystalsDeep2021, fungInverseDesignTwodimensional2021} or its underlying problem\cite{
court3DInorganicCrystal2020,hoffmann2019crystalvae,kimGenerativeAdversarialNetworks2020}, that is, the estimation of crystal structures under given conditions.  Regarding MI and ML, whether to input or output crystal structures (\ie, \emph{encode} or \emph{decode} crystal structures in MI and ML terms) induces a crucial difference. Although encoding crystal structures is suitably established using graph neural networks\cite{xieCrystalGraphConvolutional2018, chenGraphNetworksUniversal2019, parkDevelopingImprovedCrystal2020, chengGeometricinformationenhancedCrystalGraph2021}, a technical bottleneck persists for decoding crystal structures. We addressed the bottleneck in this study.

The crystal structure of an inorganic material is a regular and periodic arrangement of atoms in a three-dimensional (3D) space. This arrangement is usually described by the 3D positions and species of atoms in a unit cell and the lattice constants defining the translations of the unit cell in 3D space. The atoms in a unit cell have no explicit order, and their quantity varies from one to hundreds in number. Because ML models, including neural networks, generally accept fixed-dimensional and consistently ordered tensors for processing, treating crystal structures with ML models is not straightforward\cite{xieCrystalGraphConvolutional2018}, and determining crystal structures via the models is even more difficult.

\begin{figure}[t]
    \centering
    \includegraphics[width=14cm]{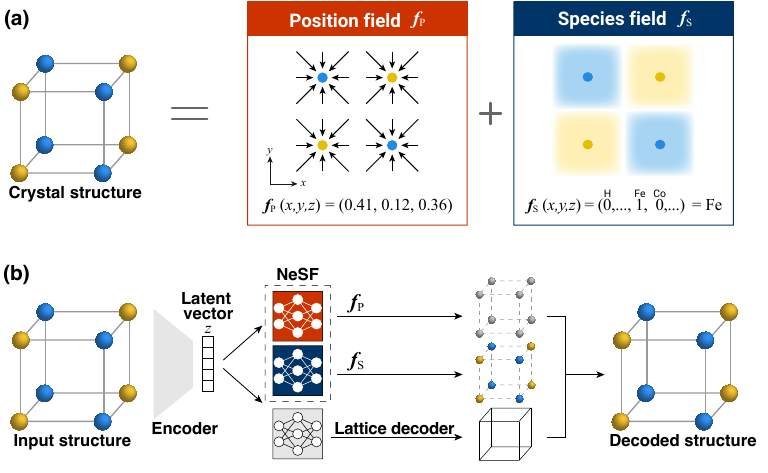}
    \caption{\textbf{Overview of the structure field and crystal structure autoencoder using the NeSF.} \textbf{a}, The structure field consists of two vector fields, namely, position field $\bm{f}_\text{p}$ and species field $\bm{f}_\text{s}$, which are defined in 3D space. Given a 3D point as a query, the position field is trained to represent the 3D vector pointing to the nearest atom of the query. In addition, the species field is trained to represent the species of the nearest atom as a categorical probability distribution.
    \textbf{b}, Network architecture of proposed crystal structure autoencoder using the NeSF. The input crystal structure is transformed into latent vector $\bm{z}$ via a PointNet-based encoder\cite{qi2017pointnet,zaheer2017deepsets}. In the decoder using the NeSF, the position field and species field are queried to recover atomic positions and species, respectively, from material information given as $\bm{z}$. The lattice constants are also estimated via the lattice decoder. These encoder and decoder are implemented as simple MLPs. See the Appendix for detailed definitions.
    \label{fig:NeSF_schematic}}
\end{figure}

We propose a general representation of crystal structures that enables neural networks to decode or determine such structures. The key concept underlying our approach is illustrated in Fig.~\ref{fig:NeSF_schematic}, where a crystal structure is represented as a continuous vector field tied to 3D space rather than as a discrete set of atoms. We refer to our approach as \emph{neural structure fields} (NeSF). The NeSF uses two types of vector fields, namely, the position and species fields, to implicitly represent the positions and species of the atoms in the unit cell of the crystal structure, respectively. 

To illustrate the concept of the NeSF, assume that we are given target material information as a fixed-dimensional vector, $\bm{z}$, and consider the problem of recovering a crystal structure from $\bm{z}$. Input $\bm{z}$ may specify, for example, information on the crystal structure of a material or some desired criteria for materials to be produced. In the NeSF, we use neural network $\bm{f}$ as an implicit function to indirectly represent the crystal structure embedded in $\bm{z}$ instead of letting neural network $\bm{f}$ directly output the crystal structure as $\bm{f}(\bm{z})$. Specifically, we treat $\bm{f}$ as a vector field on 3D Cartesian coordinates, $\bm{p}$, conditioned on the target material information, $\bm{z}$:
\begin{equation}
      \bm{s} = \bm{f}(\bm{p}, \bm{z})
\end{equation}
In the position field, network $\bm{f}$ is trained to output a 3D vector pointing from query point $\bm{p}$ to its nearest atom position, $\bm{a}$, in the crystal structure of interest. Thus, we expect output $\bm{s}$ to be $\bm{a} - \bm{p}$. If the position field is ideally trained, we can retrieve position $\bm{a}$ of the nearest atom at any query point $\bm{p}$ as $\bm{p} + \bm{f}(\bm{p}, \bm{z})$. Mathematically, the position field can be interpreted as the gradient vector field, $-\nabla \phi (\bm{p})$, of scalar potential $\phi(\bm{p}) = \frac{1}{2}\mathrm{min}_i|\bm{p} - \bm{a}_{i}|^2$, which represents the squared distance to the nearest atom given atomic positions $\{\bm{a}_{i}\}$ in the crystal structure. Analogously, the species field is trained to output a categorical probability distribution that indicates the species of the nearest atom. Thus, the output dimension of the species field is the number of candidate atomic species.

The proposed NeSF is inspired by the concepts of vector fields in classical physics and implicit neural representations\cite{mildenhall2020nerf,park2019deepsdf,chen2019imnet,mescheder2019occ,xie2022nfreview} in computer vision. Implicit neural representations have recently been proposed to handle some representation issues in 3D computer vision applications, such as 3D shape estimation of objects\cite{park2019deepsdf,chen2019imnet,mescheder2019occ} and free-viewpoint image synthesis\cite{mildenhall2020nerf,xie2022nfreview}. In 3D shape estimation, letting a neural network directly output a 3D mesh or point cloud suffers from representation issues similar to those occurring in crystal structures. To overcome these issues, the signed distance function (SDF) is utilized in DeepSDF\cite{park2019deepsdf} to model 3D shapes by letting neural network $f(\bm{p})$ indicate whether query point $\bm{p}$ is outside or inside the object volume with a positive or negative sign in its scalar outputs, respectively. The NeSF follows the basic idea of implicit neural representations and further extends it to the estimation of crystal structures described by atomic positions and species. The precise description of atomic positions in crystal structures is of crucial interest in materials science. Thus, the NeSF outputs vectors pointing  to the nearest atoms to represent atomic positions more directly than existing implicit neural representations of 3D geometries\cite{park2019deepsdf,chen2019imnet,mescheder2019occ}.

Our idea of representing crystal structures as continuous vector fields has been partially and implicitly explored using grid-based discretization (\ie, voxelization) in recent MI studies\cite{nohInverseDesignSolidState2019, nohMachineenabledInverseDesign2020a, kimGenerativeAdversarialNetworks2020, longConstrainedCrystalsDeep2021,
court3DInorganicCrystal2020,hoffmann2019crystalvae}, but without explicit consideration as discretized vector fields. In those studies, the 3D space within the unit cell is discretized into voxels, and  each voxel is then assigned an electron density, which essentially represents the presence or absence of an atom around the voxel.
However, compared with one-dimensional (\eg, audio signals) or two-dimensional (\eg, images) data, the discretization of 3D data considerably suffers from the tradeoff between spatial resolution and computational complexity in terms of both computational time and memory space. For example, the ICSG3D method\cite{court3DInorganicCrystal2020} uses $32\times32\times32$ voxels to represent crystal structures and estimates them using 3D convolutional neural networks (CNNs). Because voxel-based 3D CNNs are computationally and memory intensive, the resolution of $32\times32\times32$ voxels is an approximate limit for training a voxel-based model on a standard computing system. Meanwhile, existing crystal structures can contain tens or more atoms in their unit cells or have elongated or distorted unit cells. Thus, accurately representing diverse crystal structures with voxels requires a sufficiently high resolution. Moreover, voxel-based models can only provide atomic positions indirectly in representations such as peaks in a scalar field of electron densities discretized in the voxel space.

The proposed NeSF overcomes the limitations of voxelization. In the NeSF, there is essentially no tradeoff between the spatial resolution and required memory. Theoretically, the NeSF can achieve infinitely high spatial resolution with compact (memory- and parameter-efficient) neural networks in place of costly 3D CNNs. In addition, the NeSF can effectively represent arbitrary crystal structures including those with elongated or distorted unit cells. Furthermore, the NeSF can directly provide the Cartesian coordinates of the atomic positions rather than peaks in a scalar field. We believe that the proposed NeSF will break through a technical bottleneck in MI approaches for crystal structure estimation and contribute to the advancement of MI research in this direction.

In the following section, we detail the proposed NeSF and demonstrate its expressive power for various crystal structures through numerical experiments. Notably, the NeSF successfully recovers various crystal structures, from the relatively basic structures of perovskite materials to the complex structures of cuprate superconductors. Results from extensive quantitative evaluations demonstrate that the NeSF outperforms the voxelization approach in ICSG3D\cite{court3DInorganicCrystal2020}.

\section{Results and Discussion}
We first describe the procedures for estimating crystal structures with the NeSF and for training the NeSF. We then present an autoencoder of crystal structures as an application of the NeSF. In this autoencoder, crystal structures are embedded into vectors $\bm{z}$ (called \emph{latent vectors}) via an encoder, and then the NeSF acts as a decoder to recover the input crystal structures from $\bm{z}$. The performance of the NeSF-based autoencoder is quantitatively analyzed by evaluating the reconstruction accuracy, outperforming the voxelization-based ICSG3D baseline. 
Furthermore, we qualitatively analyze the space of vectors $\bm{z}$ learned by the proposed autoencoder. This analysis shows that the learned space reflects some similarity between crystal structures instead of merely embedding crystal structures at random.

\subsection{Crystal structure estimation with NeSF}
\label{sec:alg}

Given the target material information as vector $\bm{z}$, the estimation of the crystal structure from $\bm{z}$ amounts to estimating the positions and species of the atoms in the unit cell along with the lattice constants. 
The lattice constants are modeled as 
lengths $a$, $b$, and $c$ and angles $\alpha$, $\beta$, and $\gamma$, and estimated via simple multilayer perceptrons (MLPs) with input $\bm{z}$.
On the other hand, the atomic positions and species are estimated by position field $\bm{f}_\text{p}$ and species field $\bm{f}_\text{s}$ of the NeSF, respectively, as described in the previous section. These fields are also implemented as simple MLPs, each taking query position $\bm{p}$ and vector $\bm{z}$ as input and predicting a field value (\ie, 3D pointing vector or categorical probability distribution). An overview of the NeSF network architecture is illustrated in the right part of Fig.~\ref{fig:NeSF_schematic}\textbf{b}.

Given vector $\bm{z}$ from the encoder, the estimation of the atomic positions and species using the NeSF is illustrated in Fig.~\ref{fig:NeSF_algorithm} and summarized in the five steps:

\begin{figure*}[t]
\centering
\small
      \includegraphics[width=18cm]{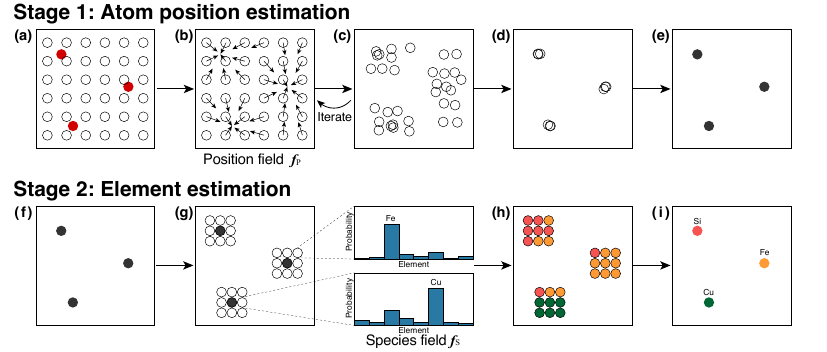}
      \caption{\textbf{Algorithm for crystal structure estimation with the NeSF.} 
      The algorithm consists of two stages: 1) atomic position estimation (corresponding to steps 1--4 in Section~\ref{sec:alg}) and 2) element estimation (corresponding to step 5 in Section~\ref{sec:alg}).
      \textbf{a}, To detect true atomic positions (shown as red dots), we regularly place query particles in the position field. 
      \textbf{b}, \textbf{c},  Then, we iteratively infer the vectors pointing to their nearest atoms and update the query particles until convergence. \textbf{d}, \textbf{e}, We detect atoms by clustering the query particles.
      \textbf{f}, \textbf{g}, For element estimation, we spread new query particles around each detected atom.
      \textbf{h}, Then, we query the species field to obtain categorical distributions as votes for the atomic species. \textbf{i}, We finally apply majority voting to determine the species of each atom.
      }
      \label{fig:NeSF_algorithm}
\end{figure*}

\begin{enumerate}
    \item \textbf{Initialize particles.}
    We first estimate the lattice constants via MLPs. Then, we regularly spread initial query points $\{\bm{p}_i^0\}$, which we call \emph{particles}, at 3D grid points within a bounding box. The bounding box is common to each dataset and is given to loosely encompass the atoms of all training samples.

\item \textbf{Move particles.}
We update the position of each particle $\bm{p}_i^t$ using the position field according to $\bm{p}_i^{t+1} = \bm{p}_i^t + \bm{f}_\text{p}(\bm{p}_i^t, \bm{z})$. We iterate this process for all particles to obtain $\{\bm{p}_i\}$ as candidate atomic positions. Because the position field is expected to point to the nearest atom position, the particles travel toward their nearest atoms through this process.

\item \textbf{Score particles.}
We score each particle $\bm{p}_i$ and filter outliers. As the norm of the output of the position field, $\| \bm{f}_\text{p}(\bm{p}_i,\bm{z}) \|$,  indicates an estimated distance from $\bm{p}_i$ to its nearest atom, we score each particle $\bm{p}_i$ by $\| \bm{f}_\text{p}(\bm{p}_i,\bm{z}) \|$ and discard the particle if the score is above a certain threshold (set to 0.9 {\AA} in this study).

\item \textbf{Detect atoms.} Until this point, the particles are expected to form clusters around atoms. Thus, we apply a simple clustering algorithm to detect each cluster position as an atomic position, determining the number of atoms in the crystal structure. Here we employ a well-known clustering algorithm in object detection, non-max suppression.
Specifically, we initialize a list of candidate particles as $B_\text{c} = \{\bm{p}_i\}$ and another list of accepted particles as $B_\text{a} = \{\}$. 1) We select the particle with the lowest score (\ie, estimated closest to an atom) from $B_\text{c}$ and move it to $B_\text{a}$. 2) We remove particles  from $B_\text{c}$ that are within a spherical area around the selected particle. (Here, we used the spherical radius of 0.5 {\AA}, and if the number of particles in the sphere is less than 10, we reject the candidate.) 
By repeating these steps until $B_\text{c}$ becomes empty, we obtain atomic positions $\{ \bm{a}_i \}$ as the selected particles stored in $B_\text{a}$.

\item \textbf{Estimate species.} Finally, we use species field $\bm{f}_\text{s}(\bm{p},\bm{z})$ to estimate the atomic species at each atomic position $\bm{a}_i$. For robust estimation against errors in $\bm{a}_i$, we spread new particles intensively around each $\bm{a}_i$ as queries to the species field instead of using $\bm{a}_i$ directly as a query. These query particles are sampled using a local regular grid centered at $\bm{a}_i$.
Hence, we obtain multiple probability distributions, each predicting the species of atom~$\bm{a}_i$. We select the most frequent atomic species among them as the final estimate.

\end{enumerate}
Note that the presented algorithm runs deterministically without any random process involved.
The hyperparameters used for sampling points in crystal structure estimation are provided in Supplementary Note 6. Sampling parameters in the Supplementary Information (SI). We also analyzed the sensitivity of these hyperparameters in Supplementary Note 3. Hyperparameters for crystal structure estimation in the SI.

\subsection{Training of NeSF}
The training of the NeSF differs from the above estimation algorithm and is much simpler. 
Specifically, we train the NeSF to directly predict the vectors pointing to the nearest atoms and categorical distributions indicating their species, without iterative position update. To do so, we randomly sample 3D query points $\{\bm{p}_i^s\}$ in the unit cell and compute loss values for the field outputs at these points, thus supervising $\bm{f}_\text{p}(\bm{p}_i^s, \bm{z})$ and $\bm{f}_\text{s}(\bm{p}_i^s, \bm{z})$ to indicate the position and species of the nearest atom, respectively. However, we cannot densely sample the query points because of practical limitations in memory usage. Therefore, a sampling strategy for query points is required for training.
Existing implicit neural representations for 3D shape estimation, such as DeepSDF\cite{park2019deepsdf}, sample the training query points near the surface. Curriculum DeepSDF\cite{curriculmdeepsdf} further introduces curriculum learning, in which the sampling density intensifies near the surface as training proceeds.

To consider a desired sampling strategy for training the position and species fields, we consider their dynamics in the proposed algorithm.
1) Particles iteratively move in the position field toward their nearest atoms. Thus, the position field should be sufficiently accurate everywhere to allow the flow of particles to their destinations and highly accurate in the vicinity of atoms.
2) The species field is queried only around atoms. Thus, it does not need to be accurate everywhere but should be robust to errors in the estimated atomic positions.

To meet the above mentioned requirements, we introduce two sampling methods to train the position and species fields.
1) Global grid sampling: This method considers 3D grid points that uniformly cover the entire unit cell and samples the points with perturbations that follow a Gaussian distribution.
2) Local grid sampling: This method considers local 3D grid points centered at each atomic position and samples the points with perturbations that follow a Gaussian distribution.

To train the position field, we combine both sampling methods. Hence, query points are sampled uniformly over the entire unit cell and densely around the atoms.
To train the species field, we use local grid sampling to concentrate training query points in the neighborhood of atoms.
The parameters used for sampling are provided in Supplementary Note 6. Sampling parameters in the SI.

\subsection{Crystal structure autoencoder}

To demonstrate and evaluate the expressive power of the NeSF, we propose an autoencoder of crystal structures.
Similar to other common autoencoders, the proposed NeSF-based autoencoder consists of an encoder and decoder. The encoder is a neural network that transforms an input crystal structure (\ie, positions and species of the atoms in the unit cell and lattice constants) into abstract latent vector $\bm{z}$. The decoder, for which we use the NeSF, reconstructs the input crystal structure from latent vector $\bm{z}$. Autoencoders are typically used to learn latent vector representations of data via self-supervised learning, in which the input data can supervise the learning via a reconstruction loss.

While we focus on decoding crystal structures, their encoding has been studied in MI. 
Because a crystal structure is essentially a set of atoms, its encoding must handle a variable number of atoms with invariance to permutation. In ML, such encoders are generally called \emph{set functions}\cite{qi2017pointnet,zaheer2017deepsets}. Among them, the family of graph neural networks\cite{xieCrystalGraphConvolutional2018, chenGraphNetworksUniversal2019, parkDevelopingImprovedCrystal2020, chengGeometricinformationenhancedCrystalGraph2021} serves as popular crystal-structure encoders. However, these networks implicitly represent atomic positions as edges, encoding distances between atoms while discarding the exact coordinates. Although this distance-based graph representation is key to ensure the invariance to coordinate systems, its information loss in input may unintentionally hinder the reconstruction performance.
Thus, we adopt the basic encoder architecture from PointNet\cite{qi2017pointnet} and DeepSets\cite{zaheer2017deepsets}.
This architecture not only represents the simplest type of set-function-based networks but can also preserve the information of input crystal structures, thus being appropriate for the performance evaluation of the NeSF decoder.
The architecture of the proposed autoencoder is detailed in Section~\ref{sec:arch} and in Supplementary Note 5. Network architecture in the SI. 

\subsubsection*{Training and evaluation procedures}
We trained and evaluated the autoencoder and ICSG3D~\cite{court3DInorganicCrystal2020} (baseline) on three materials datasets: ICSG3D, limited cell size 6 {\AA} (LCS6{\AA}), and YBCO-like datasets. These datasets collect the crystal structures of materials from the Materials Project\cite{materialsproject} and are designed to have different difficulty levels. The ICSG3D dataset is a materials collection used by Court~et al.~\cite{court3DInorganicCrystal2020} and contains 7897 materials with limited crystal systems (\ie, cubic) and prototypes  (\ie, AB, ABX2, and ABX3), and it is intended to be the easiest among the three datasets. The LCS6{\AA} dataset consists of 6005 materials with unit cell sizes of 6 {\AA} or less along the $x$, $y$, and $z$ axes and without restrictions on the crystal systems and prototypes. The YBCO-like dataset consists of 100 materials with narrow unit cells along the $c$ axis. These structures typically include those of yttrium barium copper oxide (YBCO) superconductors. Owing to the complexity of the structures and relatively few samples, the YBCO-like dataset is the most challenging among the three evaluated datasets. Further details on these datasets are provided in Section~\ref{sec:dataset}. 

For training and evaluation, we randomly split each dataset into training (90.25\%), validation (4.75\%), and test (5\%) sets. The training set was only used to train the ML models. The validation set was used to preliminarily validate the trained ML model, and the test set was used to compute the final evaluation scores after training, validation, and hyperparameter tuning. 
The hyperparameters were tuned based on the validation scores from the LCS6{\AA} dataset.
To reduce the performance variation owing to randomness (\eg, randomness in initialization of network weights), we repeated training and evaluation 10 times with different random seeds and evaluated the performance using the mean and standard deviation of the scores. 
Because the YBCO-like dataset has only 100 samples, it was treated slightly differently from the other two evaluated datasets. To reduce the performance variation owing to data splitting, we adopted twentyfold cross-validation for the YBCO-like dataset and performed one trial (instead of repeating 10 times with different random seeds).
The iterative training of the neural networks was conducted using stochastic gradient descent with Adam\cite{Adam} as the optimizer. Detailed training procedures, including the loss function definition, are provided in Section~\ref{sec:training}. 

The reconstruction performance was measured in terms of errors in the number of atoms, position, and species.
The error in the number of atoms is the rate of materials for which the number of atoms in the unit cell is incorrectly estimated.
The position error is the average error of the reconstructed atomic positions.
Depending on the denominator of the metric, we evaluated the position error in two ways. An \emph{actual} metric was used to evaluate the mean position errors at the actual atomic sites of the crystal structure by computing their shortest distances to estimated atomic sites. By contrast, a \emph{detected} metric was used to evaluate the errors at the estimated sites by computing their shortest distances to the actual atomic sites. Formally, given sets of actual and detected atomic positions as $P_\text{actual}$ and $P_\text{detected}$, the actual and detected metrics are provided as $m(P_\text{actual},P_\text{detected})$ and $m(P_\text{detected},P_\text{actual})$, respectively, with function
$m(A,B) = \sum_{\bm{a}\in A} \min_{\bm{b}\in B}\|\bm{a} - \bm{b}\|$/|A|.
The actual metric is more sensitive to errors related to underestimation of the number of atoms, whereas the detected metric is more sensitive to errors related to overestimation.
The species error is the average rate of atoms with incorrectly estimated species. Analogous to the position error, the species error was evaluated using actual and detected metrics. 
Lower values of these metrics indicate better performance.

\subsubsection*{Quantitative performance comparisons with ICSG3D}
\begin{figure*}[t]
\centering
\includegraphics[width=0.9\textwidth]{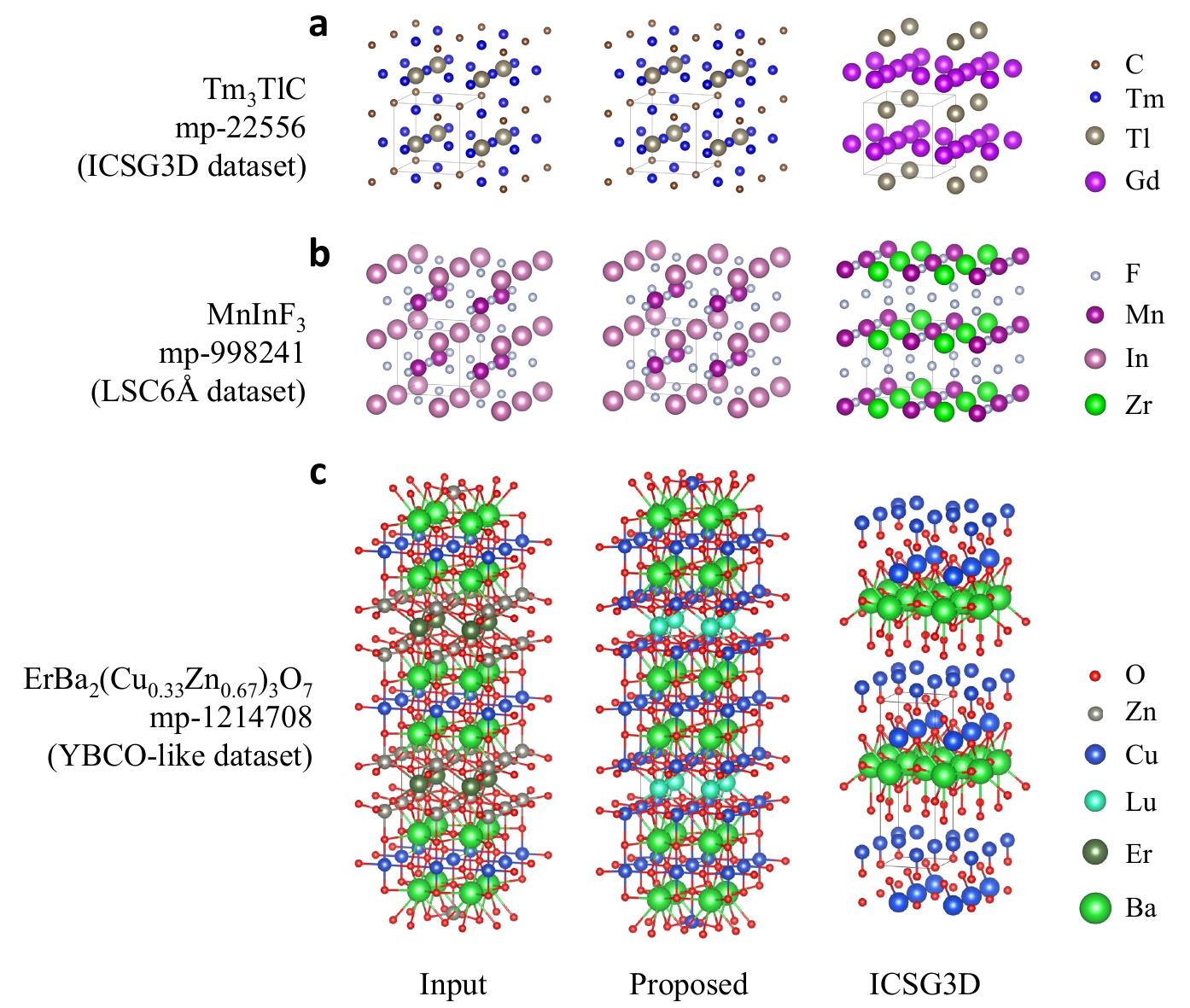}
\caption{\textbf{Reconstruction examples.} For the three evaluated datasets (rows), we show test samples (first column) and reconstruction results by the proposed (second column) and ICSG3D (third column) methods.}
\label{fig:reconstruction_samples}
\end{figure*}

Table~\ref{table:main} lists the reconstruction errors of the proposed NeSF-based autoencoder and ICSG3D baseline on the test sets of the three datasets. We also provide the scores for the training and validation sets in Supplementary Note 1. Reconstruction results for the training and validation splits in the SI.
Overall, the proposed method consistently outperforms ICSG3D in all the evaluation metrics, with substantial performance improvements for the species error in all datasets and for all the metrics on the YBCO-like dataset. Fig.~\ref{fig:reconstruction_samples} shows crystal structures from the three evaluated datasets, comparing test samples and reconstruction results by the proposed autoencoder and ICSG3D.

\begin{table}[btp]
\centering
\caption{\textbf{Reconstruction results (test set).}}
\label{table:main}
\scalebox{0.725}[0.725]{
\begin{tabular}{lrrrrrr}
\toprule
Method & \multicolumn{1}{c}{Proposed} & \multicolumn{1}{c}{ICSG3D} & \multicolumn{1}{c}{Proposed} & \multicolumn{1}{c}{ICSG3D} & \multicolumn{1}{c}{Proposed} & \multicolumn{1}{c}{ICSG3D} \\
\midrule
Dataset & \multicolumn{2}{c}{ICSG3D} & \multicolumn{2}{c}{LCS6\AA} & \multicolumn{2}{c}{YBCO-like} \\
\midrule
Error in number of atoms [\%] & $0.53\pm0.25$ & $2.67\pm0.84$ & $6.35\pm0.50$ & $17.28\pm1.13$ & $12.00$ & $91.00$ \\
Position error (actual) [\AA] & $0.0308\pm0.0112$ & $0.0877\pm0.0306$ & $0.1161\pm0.0073$ &$0.2006\pm0.0132$  & $0.2631$&  (0.6311) \\
Position error (detected) [\AA] & $0.0359\pm0.0226$ & $0.1057\pm0.0284$ & $0.1632\pm0.0123$ & $0.2886\pm0.0159$  & $0.2448$ & $0.4358$\\
Species error (actual) [\%] & $4.31\pm0.39$ & $64.39\pm1.91$ & $14.78\pm0.81$ & $55.70\pm1.40$ & $14.78$ & $78.08$\\
Species error (detected) [\%] & $4.36\pm0.39$ & $65.05\pm1.85$ & $16.11\pm0.66$ & $58.32\pm1.45$ & $19.89$& $54.24$ \\
Lattice length error [\AA] & $0.27 \pm 0.74$ & $0.08 \pm 0.04$ & $0.05 \pm 0.00$ & $0.06 \pm 0.01$ & $0.25$ & $0.10$ \\
Lattice angle error [deg] & $0.37 \pm 1.10$ & $0.00 \pm 0.00$ & $0.19 \pm 0.04$ & $0.34 \pm 0.07$ & $2.91$ & $0.06$ \\
\bottomrule
\end{tabular}
}\\
\vskip 5pt
\begin{minipage}[c]{0.9\textwidth}
\footnotesize
We compared the performance of the proposed NeSF-based crystal structure autoencoder and ICSG3D method for crystal structure reconstruction on three datasets (ICSG3D, LCS6{\AA}, and YBCO-like). The position and species errors were evaluated in two ways based on the actual or detected atomic sites. For the YBCO-like dataset, ICSG3D failed to output any atom for 44 out of the 100 materials and provided no valid score for the position error (actual). We computed a score by excluding those incorrectly estimated materials (shown in parentheses), indicating that the actual performance of ICSG3D is worse the displayed value. 
\end{minipage}
\end{table}

\begin{figure*}[htbp]
\centering
\begin{overpic}[width=0.45\linewidth]{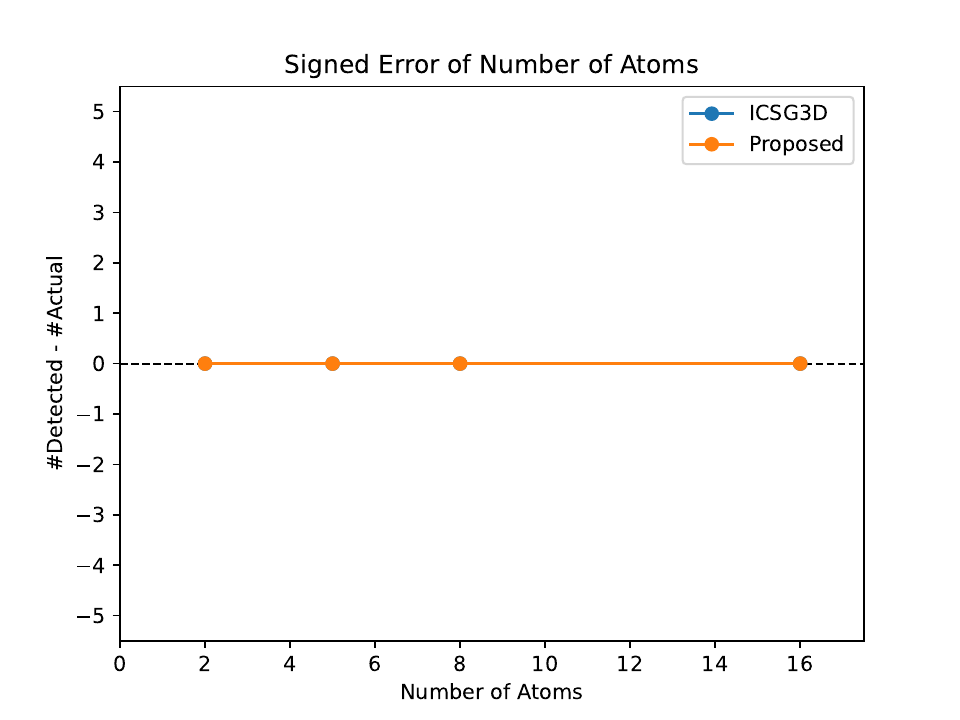}
\put(5,155){\textbf{(a)}}
\end{overpic} \hfil
\begin{overpic}[width=0.45\linewidth]{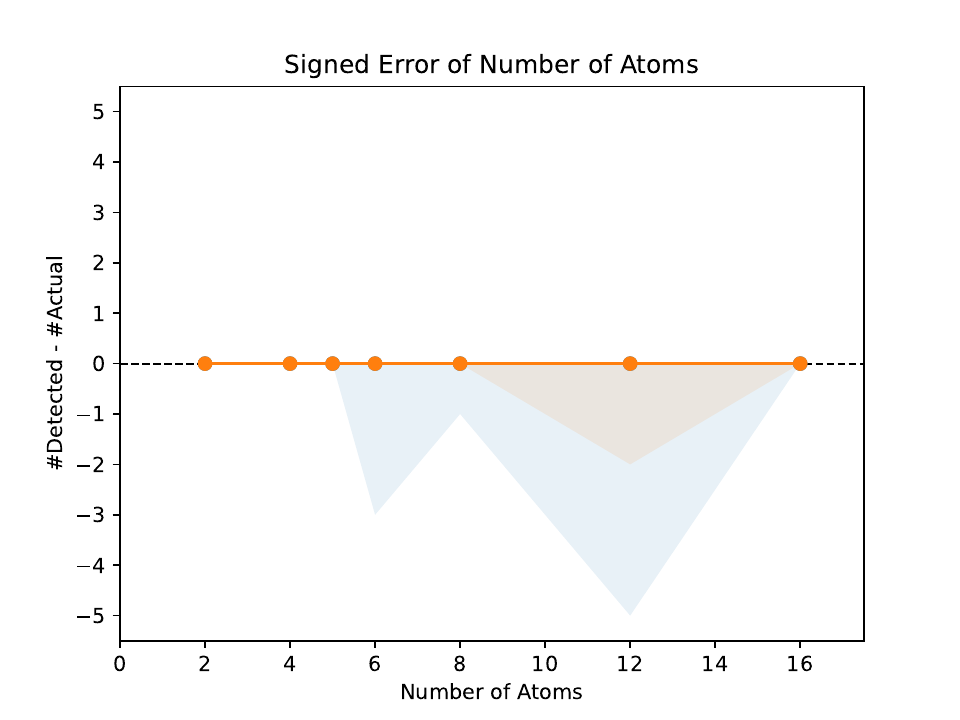}
\put(5,155){\textbf{(b)}}
\end{overpic}\\
\begin{overpic}[width=0.45\linewidth]{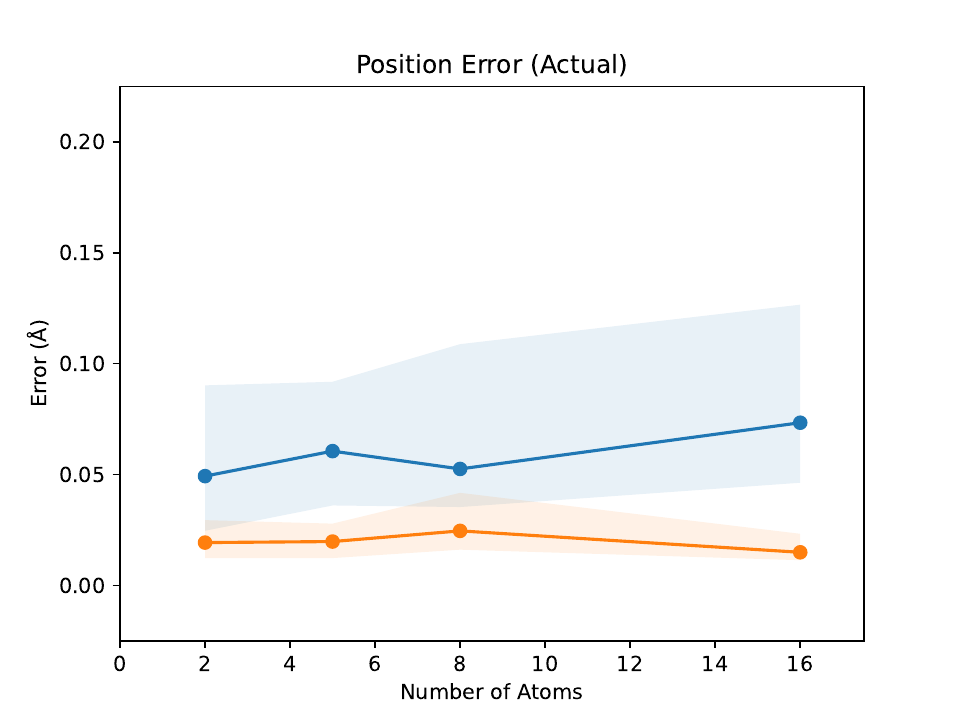}
\put(5,155){\textbf{(c)}}
\end{overpic}\hfil
\begin{overpic}[width=0.45\linewidth]{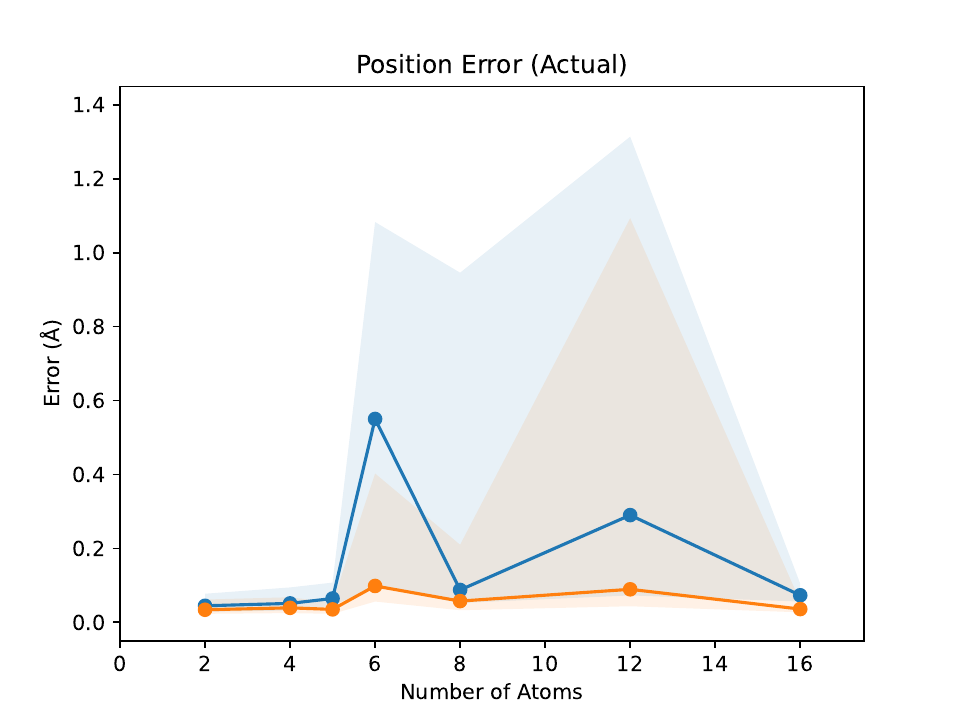}
\put(5,155){\textbf{(d)}}
\end{overpic}\\
\begin{overpic}[width=0.45\linewidth]{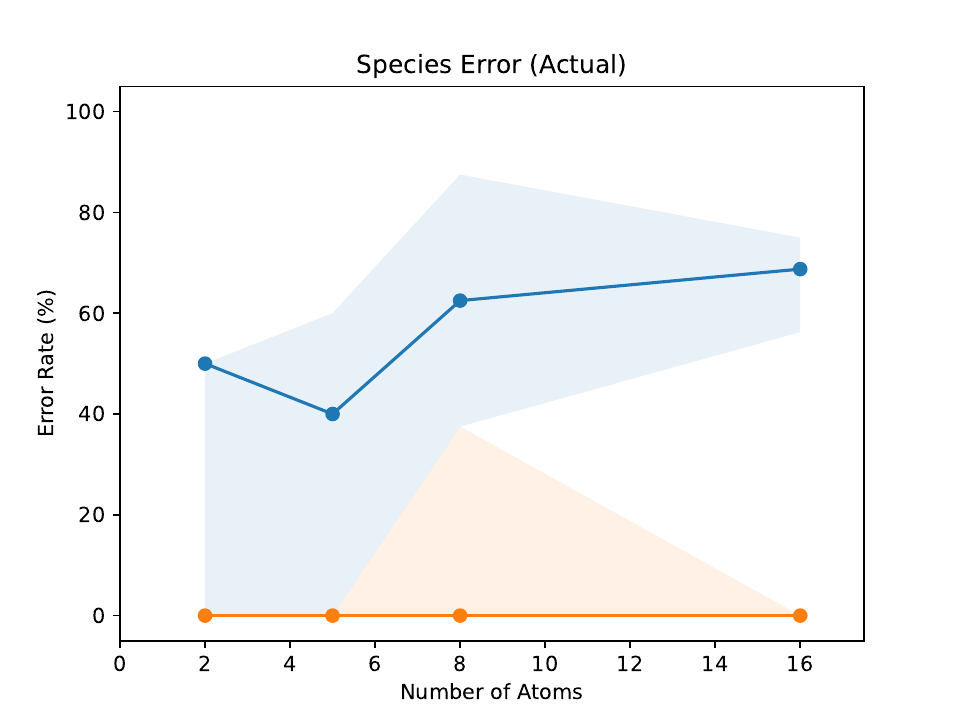}
\put(5,155){\textbf{(e)}}
\end{overpic} \hfil
\begin{overpic}[width=0.45\linewidth]{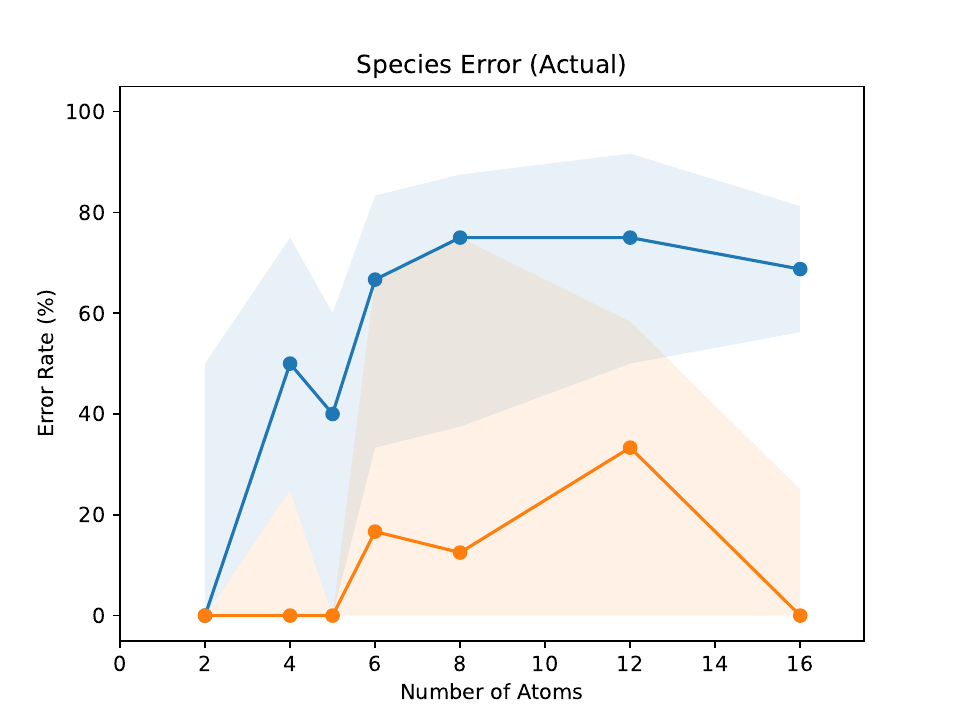}
\put(5,155){\textbf{(f)}}
\end{overpic}\\
\begin{minipage}{0.45\linewidth}
\centering
ICSG3D dataset
\end{minipage} \hfil
\begin{minipage}{0.45\linewidth}
\centering
LCS6{\AA} dataset
\end{minipage}
  \caption{\textbf{Relationships between reconstruction performance and number of atoms in conventional unit cell.} Three kinds of error distributions (rows) are shown for the ICSG3D (first column) and LCS6{\AA} (second column) datasets described by medians (points) and 68-percent ranges (colored regions) similar to means $\pm \sigma$ ranges. \textbf{a, b}, Distributions of signed errors between detected and actual numbers of atoms. \textbf{c, d}, Distributions of position errors according to number of atoms averaged over individual structures. \textbf{e, f}, Distributions of species errors according to number of atoms averaged over individual structures. For statistical reliability, we omit the $x$-axis points counting less than 10 materials samples from the visualization (see Section~\ref{sec:dataset} for detailed dataset statistics).}
      \label{fig:plot}
\end{figure*}

For the ICSG3D dataset, which is the easiest of the three datasets, ICSG3D achieves good performance for the error in number of atoms and position error, but provides high species error (approximately 65\% in both actual and detected metrics). By contrast, the proposed method achieves slightly better position error and error in number of atoms and drastically lower species error (approximately 4\%).
We believe that this is because ICSG3D estimates the atomic species via electron density maps, whereas the proposed method more directly represents atomic species as categorical distributions. Extending ICSG3D to estimate a categorical distribution at each voxel is impractical because it would require approximately $100\times 32^3$ times the memory usage in the output (\ie, requiring 100 species categories for each of the $32^3$ voxels in addition to one electron density map).

For the the LCS6{\AA} and YBCO-like datasets, which are more challenging than the ICSG3D dataset, 
the performance advantage of the proposed method is even clearer, especially regarding the position error.
The LCS6{\AA} dataset contains a variety of crystal structures (\eg, non-cubic structures, distorted crystal structures), whereas the YBCO-like dataset contains very narrow crystal structures.
In addition, the YBCO-like dataset contains few samples, possibly leading to model overfitting (\ie, the performance on the test set is likely to degrade significantly).
Despite these difficulties, the proposed method can accurately estimate the atomic positions and species.

For a detailed analysis of the relationship between the method performance and structural complexity, Fig.~\ref{fig:plot} shows the distributions of reconstruction errors according to number of atoms given as medians (points) and 68-percent ranges around them (colored regions) over 10 test trials on materials from the ICSG3D and LCS6{\AA} datasets. The YBCO-like dataset is excluded from this analysis because it contains only materials with 13 atoms in their unit cells. 
Figs.~\ref{fig:plot}\textbf{a} and \ref{fig:plot}\textbf{b} show the signed errors between the numbers of detected and actual atoms. These results indicate that both methods correctly estimate the number of atoms for most (\ie, $>$ 68 percent) of the samples in the ICSG3D dataset (Fig.~\ref{fig:plot}\textbf{a}). However, the ICSG3D method underestimates the number of atoms for the LCS6{\AA} dataset (Fig.~\ref{fig:plot}\textbf{b}). 
Likewise, Figs.~\ref{fig:plot}\textbf{c} and \ref{fig:plot}\textbf{d} show the distributions of position errors and Figs.~\ref{fig:plot}\textbf{e} and \ref{fig:plot}\textbf{f} show the distributions of species errors.
Because the number of atoms is either correctly estimated or underestimated by both methods in most cases, we report the errors in the actual metrics. 
Examining the distributions at $x=2$ in Fig.~\ref{fig:plot}\textbf{e}, we can determine the trends of species errors for the diatomic structures in the ICSG3D dataset. The proposed method provides the correct species of both atoms for more than 68 percent of the diatomic materials, whereas ICSG3D often misestimates the species of one of the two atoms.

Overall, although the three types of errors by both methods tend to increase with the number of atoms, our method outperforms ICSG3D consistently for materials with varying numbers of atoms. Compared with our method, the performance of ICSG3D tends to degrade more notably for materials with many atoms. Because ICSG3D often underestimates the number of atoms (Fig.~\ref{fig:plot}\textbf{b}), this overall degradation trend suggests that ICSG3D cannot capture those many-atom structures owing to its spatial resolution limited to $32\times 32 \times 32$ voxels.

We believe that the notable high performance of the proposed method is attributable to two reasons.
First, our method does not use discretization, being advantageous over the grid-based ICSG3D for estimating complex crystal structures. In a grid-based method, the spatial resolution is limited by the cubically increasing computations and memory usage. The proposed NeSF is free from such tradeoff between resolution and computational complexity, and it can thus effectively represent complex structures.
Second, the model size of the proposed NeSF using MLPs is much smaller than that of the 3D CNN architecture of ICSG3D. In general, the number of  training samples required for an ML model is correlated with the number of trainable parameters. Grid-based methods use layers of 3D convolution filters that involve many trainable parameters. By contrast, the NeSF employs implicit neural representations to indirectly describe the 3D space as a field instead of voxels. Thus, it is efficiently implemented by MLPs with fewer parameters than a 3D CNN. Specifically, the NeSF-based autoencoder has $0.76$ million parameters, which is only $2.24$\% of the number of parameters in the 3D CNN-based ICSG3D ($34$ million parameters). This difference can make the NeSF advantageous over grid-based methods, especially on small datasets such as the YBCO-like dataset.

\subsection{Latent space interpolation}
\label{sec:interp}

We qualitatively analyzed the characteristics of the proposed NeSF-based autoencoder by inspecting the learned latent space of the crystal structures. 
In general, a good latent space should map similar items (in terms of properties, characteristics, categories, etc.) closely in the space, thus providing latent data representations that facilitate analysis by humans and machines. To assess the construction of the latent space of crystal structures, we visualized transitions in the latent space as sequences of crystal structures.
If the latent space is trained to capture relationships between materials in terms of structural similarity, interpolating between two points in the latent space should produce a sequence of materials with similar crystal structures.

Interpolation analysis proceeded as follows:
\begin{enumerate}
\item Select two known crystal structures from the ICSG3D test set as the source and destination materials and obtain their latent vectors as $\bm{z}_\text{src}$ and $\bm{z}_\text{dst}$ via the trained encoder.
\item Interpolate between $\bm{z}_\text{src}$ and $\bm{z}_\text{dst}$ linearly in the latent space to obtain a sequence of latent vectors.
\item Decode each latent vector via the trained decoder (NeSF) to obtain its crystal structure.
\end{enumerate}
Because the intermediate crystal structures between the source and destination are reconstructed from the latent vectors via the trained decoder, these structures may not appear in the dataset.

To facilitate interpretation of the analysis, we chose the well-known zinc-blende and rock-salt structure families as benchmark materials.
Both families have compositions given by \ce{AX}, where A is a cation and X is an anion, and the crystal structure is based on a cubic crystal system.
Therefore, if the source and destination materials belong to one of these families, the characteristic composition and structural prototype should be preserved throughout the interpolation path.

\begin{figure*}[t]
\centering
\small
      \includegraphics[width=\textwidth]{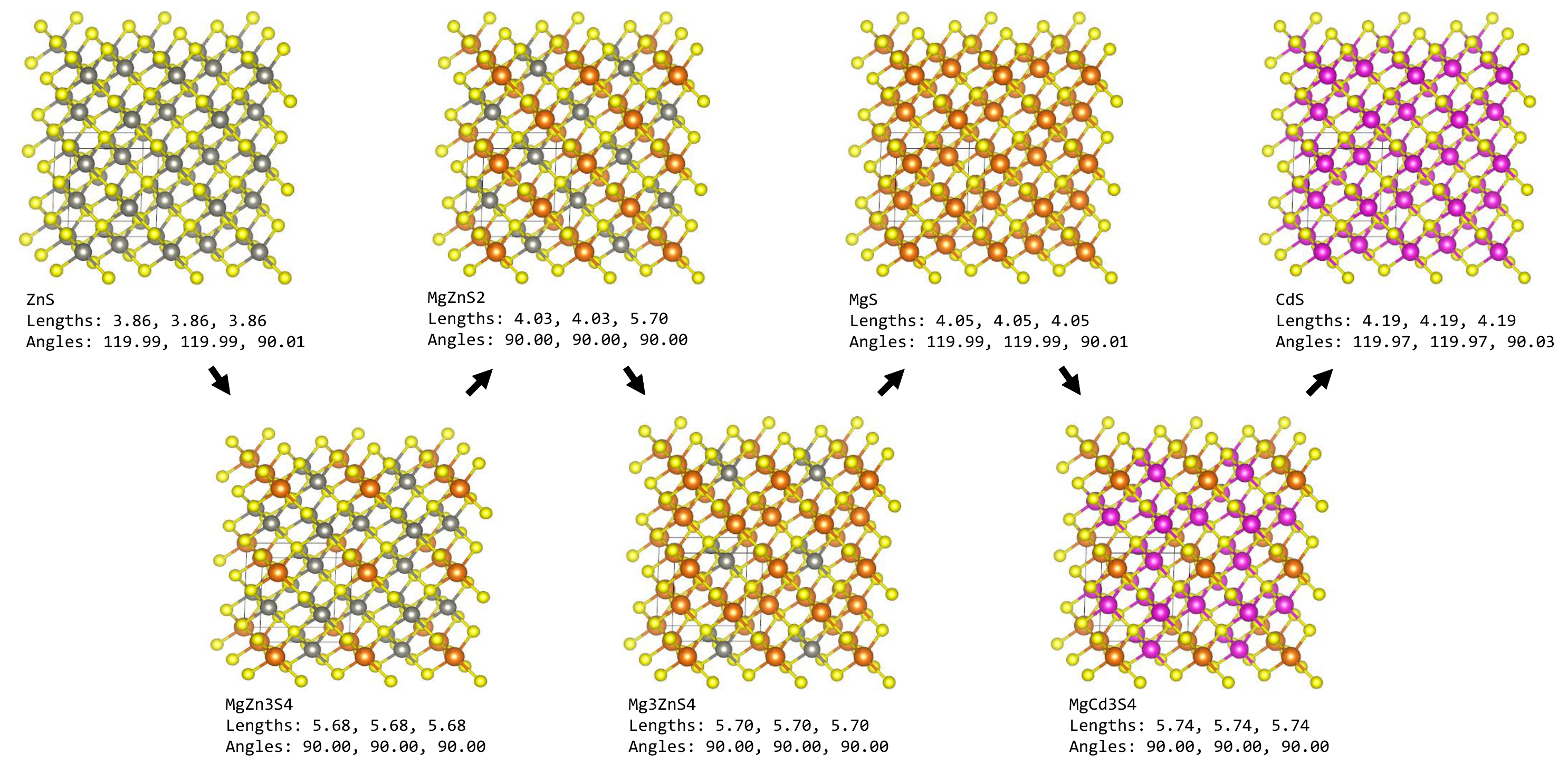}
      \caption{\textbf{Results of interpolation from \ce{ZnS} to \ce{CdS}.}}
      \label{fig:mophing_path1}
\end{figure*}
\begin{figure*}[t]
\centering
\small
      \includegraphics[width=\textwidth]{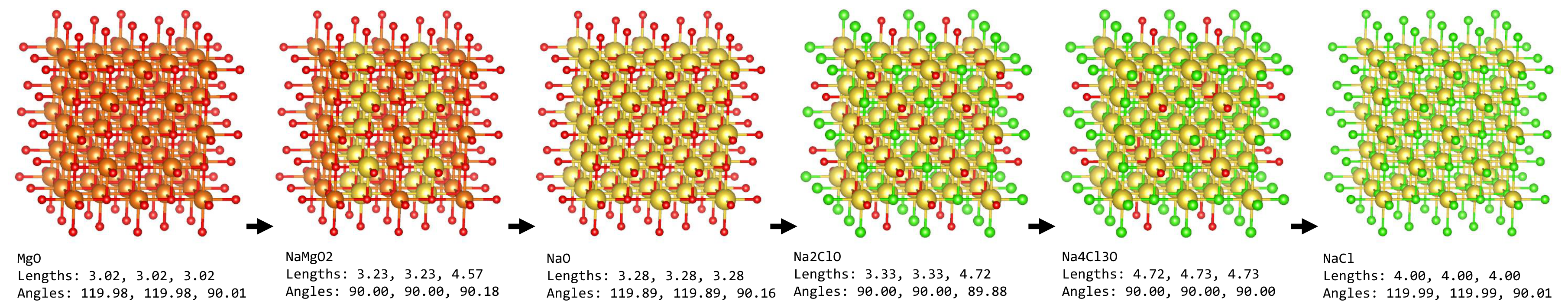}
      \caption{\textbf{Results of interpolation from \ce{MgO} to \ce{NaCl}.}}
      \label{fig:mophing_path2}
\end{figure*}

As a first example, Fig.~\ref{fig:mophing_path1} shows the results of interpolation from \ce{ZnS} (mp-10695) to \ce{CdS} (mp-2469). The obtained transition in the compositional formula is \ce{ZnS} $\to$ \ce{MgZn3S4} (\ce{Mg_{0.25}Zn_{0.75}S}) $\to$ \ce{MgZnS2} (\ce{Mg_{0.5}Zn_{0.5}S}) $\to$ \ce{Mg3ZnS4} (\ce{Mg_{0.75}Zn_{0.25}S}) $\to$ \ce{MgS} $\to$ \ce{MgCd3S4} (\ce{Mg_{0.25}Cd_{0.75}S}) $\to$ \ce{CdS}.

As a second example, Fig.~\ref{fig:mophing_path2} shows the results of interpolation from \ce{MgO} (mp-1265) to \ce{NaCl} (mp-22862). The obtained transition in the compositional formula is \ce{MgO} $\to$ \ce{NaMgO2} (\ce{Na_{0.5}Mg_{0.5}O}) $\to$ \ce{NaO} $\to$ \ce{Na2ClO} (\ce{NaCl_{0.5}O_{0.5}}) $\to$ \ce{Na4Cl3O} (\ce{NaCl_{0.75}O_{0.25}}) $\to$ \ce{NaCl}.

Additionally, we provide the results of interpolation from \ce{NaCl} (mp-22862) to \ce{PbS}, from \ce{MgO} (mp-1265) to \ce{CaO} (mp-2605), from \ce{PbS} (mp-21276) to \ce{CaO} (mp-2605), and from \ce{BaTiO3} (cubic; mp-2998) to \ce{BaTiO3} (tetragonal; mp-5986) in Supplementary Note 2. Additional results of latent space interpolation in the SI.

In the interpolation examples, composition \ce{AX} and the cubic structure are mostly preserved.
Furthermore, the compositions change continuously without collapsing along the interpolation paths. 
These results suggest that our encoder learns meaningful continuous representations of crystal structures by capturing their characteristics in an abstract space, and the proposed NeSF model successfully decodes these representations into crystal structures.

\subsection{Latent space visualization}

\begin{figure*}[tbp]
      \includegraphics[width=0.95\linewidth]{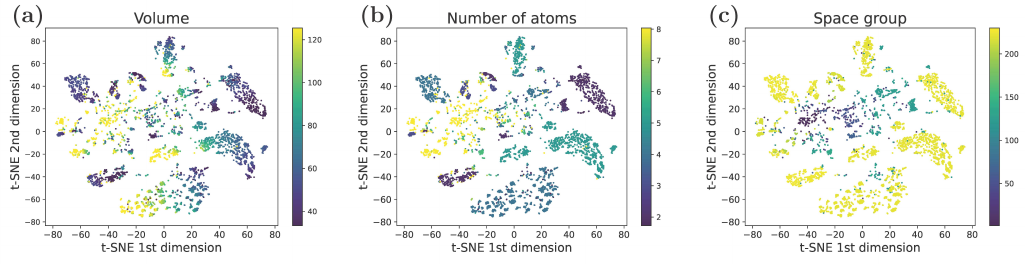}
      \caption{\textbf{t-SNE visualizations of the latent space.} Each 2D plot represents the latent vector of a material from the LCS6{\AA} dataset. The plot colors visualize (\textbf{a}) cell volume, (\textbf{b}) number of atoms in the cell, and (\textbf{c}) space group as structural attributes of the materials.}\label{fig:maps}
      \end{figure*}

As another qualitative analysis of the latent space, we visualized the space using t-distributed stochastic neighbor embedding (t-SNE)\cite{vandermaaten08a}. While the interpolation analysis in the previous section examines the local smoothness of the space, this t-SNE analysis inspects the space at a more global scale to see if it maps similar structures closely, or as clusters.
In Fig.~\ref{fig:maps}, we have embedded the 192-dimensional latent vectors of the materials from the LCS6{\AA} dataset into two-dimensional (2D) plots via t-SNE. Each plot is colored according to the cell volume (Fig.~\ref{fig:maps}a), number of atoms (Fig.~\ref{fig:maps}b), and space group (Fig.~\ref{fig:maps}c) as structural attributes of the material. 
These plots form clusters whose colors are well separated or smoothly transitioned, showing that the latent space reflects the structural similarity of the materials.

\subsection{Limitations and future directions}
This study was mainly focused on developing a fundamental approach for crystal structure estimation using implicit neural representations. To suggest room for further improvement and important directions in future work, we identified three main limitations of the NeSF.

First, the proposed NeSF does not explicitly consider space-group symmetry. Thus, the local spatial arrangements of atoms in conventional unit cells estimated by the NeSF do not necessarily obey space-group symmetry. Although ICSG3D\cite{court3DInorganicCrystal2020} shares the same limitation, symmetry is an important concept in crystallography. Thus, incorporating the constraint of space-group symmetry into the NeSF is an important direction of future work.

Second, for an evaluation purpose, we adopted the autoencoder architecture rather than generative models, such as variational autoencoders\cite{hoffmann2019crystalvae,Goodfellow_book,xieCrystalGraphConvolutional2018} and generative adversarial networks\cite{Goodfellow2014GAN,Goodfellow_book,nouiraCrystalGANLearningDiscover2019,kimGenerativeAdversarialNetworks2020}. These generative models intentionally perturb latent structural representations to produce diverse structures that do not appear in the dataset. While this aspect of the generative models is more suitable for novel structure discovery, the lack of ground-truth structures prevents quantitative and reliable performance analysis. The NeSF should be applied to generative models in future work with appropriate performance analysis.

Third, the trained latent vectors have limited capability as material descriptors for property prediction, as shown in an additional analysis provided in Supplementary Note 4. Prediction of material property in the SI. Although this result is understandable since the latent space is not directly trained for property prediction, learning a latent space that effectively disentangles material properties from structural features is another future direction. Such a latent space will be useful not only for property prediction but also for inverse design of materials with desired properties when combined with the aforementioned generative approach.

\section{Conclusion}
We propose the NeSF to estimate crystal structures using neural networks. Determining crystal structures directly using neural networks is challenging because these structures are essentially represented as an unordered set with varying numbers of atoms.
NeSF overcomes this difficulty by treating the crystal structure as a continuous vector field rather than as a discrete set of atoms. 
We borrow the idea of the NeSF from vector fields in physics and the recent implicit neural representations in computer vision. An implicit neural representations is an ML technique to represent 3D geometries using neural networks. The NeSF extends this technique by introducing the position and species fields to estimate the atomic positions and species of crystal structures, respectively. Unlike existing grid-based approaches for representing crystal structures, 
the NeSF is free from the tradeoff between spatial resolution and computational complexity and can represent any crystal structure.

The NeSF was applied as an autoencoder for crystal structures and demonstrated its performance and expressive power on datasets with diverse crystal structures. A quantitative performance analysis showed a clear advantage of the NeSF-based autoencoder over an existing grid-based method, especially for estimating complex crystal structures. Furthermore, a qualitative analysis of the learned latent space revealed that the autoencoder captures similarities between crystal structures rather than mapping crystal structures randomly.

In materials science, the design and construction of crystal structures are fundamental processes when  searching for materials with the desired properties. ML is advancing rapidly with the development of neural networks, and representing arbitrary crystal structures using those networks is essential for next-generation development of materials. For instance, the NeSF can be applied to powerful deep generative models, such as variational autoencoders and generative adversarial networks, to discover novel crystal structures. Such generative models for crystal structures will be important for inverse design of materials, which is a major challenge in MI. We believe that the NeSF can overcome the technical bottleneck of ML for crystal structure estimation and pave the way for next-generation development of materials.

\section{Methods}

\subsection{Datasets}
\label{sec:dataset}

We evaluated the proposed NeSF-based autoencoder using datasets from the Materials Project\cite{materialsproject}.
\begin{itemize}
\item ICSG3D dataset: To compare with the ICSG3D baseline, we reproduced the dataset used in the original paper\cite{court3DInorganicCrystal2020}. Unfortunately, the proponents of ICSG3D did not specify the identifiers of the materials used for their experiments. Therefore, by following the procedures described in their paper, we crawled the three material classes with cubic structures, namely, binary alloys (\ce{AB}), ternary perovskites (\ce{ABX3}), and Heusler compounds (\ce{ABX2}), from the Materials Project\cite{materialsproject}. Then, we split the dataset into training (7194 samples), validation (342 samples), and test (360 samples) sets. The numbers of atoms per unit cell are listed in Table~\ref{tab:num_atoms_icsg3d}.
\item LCS6{\AA} dataset: From the Materials Project\cite{materialsproject}, we collected all the materials samples with unit cell sizes of 6 {\AA} or less along the $x$, $y$, and $z$ axes. Unlike the ICSG3D dataset, the LCS6{\AA} dataset has no restriction on material classes and prototypes and covers $4.31$\% of the Materials Project database. This dataset was also split into training (5418 samples), validation (286 samples), and test (301 samples) sets, and the numbers of atoms in the samples are listed in Table~\ref{tab:num_atoms_lcs6}.
\item YBCO-like dataset: From the Materials Project\cite{materialsproject}, we extracted 100 samples with structures similar to YBCO (\ce{YBa2Cu3O7}), that is, narrow crystal structures along the $c$ axis. The structures of these materials have narrow and anisotropic unit cells and contain various elements. Given the complexity of the structures and limited number of samples, this dataset is considered the hardest and most practical among the three evaluated datasets. This dataset was also split into  training (90 samples), validation (5 samples), and test (5 samples) sets. 
The numbers of atoms of the samples are listed in Table~\ref{tab:num_atoms_ybco}.
\end{itemize}

All the crystal structures were described by a conventional cell for the input and output of the autoencoders.

\begin{table}[htbp]
\centering
\caption{\textbf{Numbers of atoms per unit cell in ICSG3D dataset.}}
\label{tab:num_atoms_icsg3d}
\scalebox{0.725}{
\begin{tabular}{@{}l|rrrrrrrrrr@{}}
No. atoms per cell & 2 & 5 & 6 & 8 & 12 & 15 & 16 & 20 & 32 & 40 \\ \hline
Training & 484 & 1413 & 7 & 572 & 1 & 12 & 4616 & 17 & 18 & 54 \\
Validation & 18 & 66 &  & 32 &  &  & 223 & 1 &  & 2 \\
Test & 24 & 61 &  & 34 &  &  & 237 &  &  & 4
\end{tabular}
}
\end{table}

\begin{table}[htbp]
\centering
\caption{\textbf{Numbers of atoms per unit cell in LCS6{\AA} dataset.}}
\label{tab:num_atoms_lcs6}
\scalebox{0.725}{
\begin{tabular}{@{}l|rrrrrrrrrrrrrrrrrrrrrr@{}}
No. atoms per cell & 1 & 2 & 3 & 4 & 5 & 6 & 7 & 8 & 9 & 10 & 11 & 12 & 13 & 14 & 15 & 16 & 18 & 20 & 21 & 22 & 24 & 28 \\ \hline
Training & 21 & 599 & 44 & 1411 & 1325 & 244 & 69 & 686 & 58 & 137 & 9 & 352 & 9 & 24 & 4 & 402 & 5 & 6 & 1 & 2 & 9 & 1 \\
Validation &  & 35 & 2 & 80 & 81 & 14 & 3 & 41 & 4 & 5 &  & 13 &  & 2 &  & 20 &  &  &  &  & 1 &  \\
Test &  & 30 & 3 & 83 & 68 & 17 & 2 & 35 &  & 5 &   & 19 &  & 1 &  & 23 &  &  &  &  &  & 
\end{tabular}
}
\end{table}

\begin{table}[htbp]
\centering
\caption{\textbf{Numbers of atoms per unit cell in YBCO-like dataset.}}
\label{tab:num_atoms_ybco}
\scalebox{0.725}{
\begin{tabular}{@{}l|r@{}}
No. atoms per cell & 13 \\ \hline
Training & 90 \\
Validation & 5 \\
Test & 5
\end{tabular}
}
\end{table}

\subsection{Neural network architecture}
\label{sec:arch}
The proposed NeSF-based autoencoder consists of simple MLPs. Specifically, the encoder first transforms the atomic position and species of each atom into a 512-dimensional feature vector. These feature vectors per atom are then aggregated into a single feature vector via a max pooling and converted into a 192-dimensional feature vector, which represents latent vector $\bf{z}$ of the input crystal structure. Given $\bf{z}$ as the input, the decoder estimates four types of structural information features using four separate MLPs. The atomic positions are estimated via the position field implemented as an MLP with nine fully connected layers. The atomic species are estimated via the ppecies field implemented as an MLP with three fully connected layers. 
Lattice lengths ($a, b, c$) and angles ($\alpha, \beta, \gamma$) are estimated using the corresponding MLPs with two fully connected layers. More details on the architecture are provided in Supplementary Note 5. Network architecture in the SI. 

\subsection{Training}
\label{sec:training}
During training, the four types of outputs from the decoder (\ie, atomic position, species, lattice lengths, and angles) were evaluated using the following loss function:
\begin{equation}
    L = \lambda_\text{pos} L_\text{pos} + \lambda_\text{spe} L_\text{spe} + \lambda_\text{len} L_\text{len} + \lambda_\text{ang} L_\text{ang}
\end{equation}
where $L_\text{pos}$ is the mean squared error between the estimated and true atomic positions, $L_\text{spe}$ is the cross-entropy loss function for the atomic species distributions, and $L_\text{len}$ and $L_\text{ang}$ are the mean squared error of the lattice lengths and angles, respectively. 
The total loss function, $L$, is given by the weighted sum of these loss functions with weights $(\lambda_\text{pos},\lambda_\text{spe},\lambda_\text{len},\lambda_\text{ang})=(10, 0.1, 1, 1)$.
We optimized the loss function using stochastic gradient descent with a batch size of $128$. We used Adam as the optimizer with an initial learning rate of $10^{-3}$ decaying every 640 epochs by a factor of $0.5$. For each dataset, we conducted iterative training for 3200 epochs for all the materials samples in the dataset. 
We applied early stopping based on the validation score for both the proposed method and ICSG3D, except for the YBCO-like dataset, for which the few validation samples may lead to unreliable validation scores.
The training of the NeSF model took approximately 11 hours on the ICSG3D dataset, 9 hours on the LCS6{\AA} dataset, and 1 hour on the YBCO-like dataset using a computer equipped with a single Quadro RTX8000 graphics processor. For details regarding our strategies for validating the trained models and tuning the hyperparameters (\eg, latent dimensionality, number of network layers, training batch size), see Supplementary Note 7. Hyperparameter search in the SI.

\bibliographystyle{naturemag}
\bibliography{references}

\section*{Data availability}
The dataset generation procedure is available in Supplementary Note 1. Reconstruction results for the training and validation splits in the SI. Datasets will be available online here: \url{https://github.com/omron-sinicx/neural-structure-field} and also available from the corresponding author upon reasonable request.

\section*{Code availability}
The implementation details are available Supplementary Note 5. Network architecture in the SI. Code will be available online here: \url{https://github.com/omron-sinicx/neural-structure-field} and also available from the corresponding author upon reasonable request.

\section*{Acknowledgments}
This work is partly supported by JST-Mirai Program, Grant Number JPMJMI19G1 and JST Moonshot R\&D Grant Number JPMJMS2236, and MEXT Program: Data Creation and Utilization-Type Material Research and Development Project (Digital Transformation Initiative Center for Magnetic Materials) Grant Number JPMXP1122715503.
Y.S. is supported by JST ACT-I grant number JPMJPR18UE.
We would like to thank Editage (www.editage.com) for English language editing.
Computational resource of AI Bridging Cloud Infrastructure (ABCI) provided by National Institute of Advanced Industrial Science and Technology (AIST) was partly used for comparison methods.

\section*{Author contributions}
N.C. conceived the idea for the present work, did most of the implementation, and performed the numerical experiments. Y.S. conceived the idea for the present work and performed the numerical experiments. T.T. assisted in the implementation of the data analysis and discussed the results from a machine learning perspective. R.I. reviewed the implementation and discussed the results from a materials science perspective. Y.U. initiated the idea of decoding crystal structure as point clouds, directed the project, and discussed the results from a machine learning perspective. K.S. helped with the material data handling and discussed the results from a materials science perspective. K.O. directed the project and discussed the results from a materials science perspective. All authors discussed the results and wrote the manuscript together.

\section*{Competing interests}
The authors declare no conflicts of interest associated with this manuscript.

\end{document}


\ifpreview
\maketitle
\else
\maketitle
\thispagestyle{empty}
\fi

\section*{Supplementary Note 1. Reconstruction results for the training and validation splits}

\begin{table}[btp]
\centering
\caption{\textbf{Reconstruction results (train set).}}
\label{table:main_train}
\scalebox{0.725}{
\begin{tabular}{lrrrrrr}
\toprule
Method & \multicolumn{1}{c}{Proposed} & \multicolumn{1}{c}{ICSG3D} & \multicolumn{1}{c}{Proposed} & \multicolumn{1}{c}{ICSG3D} & \multicolumn{1}{c}{Proposed} & \multicolumn{1}{c}{ICSG3D} \\
\midrule
Dataset & \multicolumn{2}{c}{ICSG3D} & \multicolumn{2}{c}{LCS6\AA} & \multicolumn{2}{c}{YBCO-like} \\
\midrule
Error in number of atoms [\%] & $0.61 \pm 0.05$ & $2.53 \pm 0.35$ & $6.71 \pm 0.61$ & $13.50 \pm 1.24$ & $0.00$ & $91.82$ \\
Position error (actual) [\AA] & $0.0308 \pm 0.0039$ & $0.1160 \pm 0.0330$ & $0.1172 \pm 0.0189$ & $0.2732 \pm 0.0145$ & $0.0532$ & $0.6155$ \\
Position error (detected) [\AA] & $0.0237 \pm 0.0041$ & $0.0910 \pm 0.0336$ & $0.0754 \pm 0.0101$ & $0.1998 \pm 0.0079$ & $0.0532 $ & $0.3680$ \\
Species error (actual) [\%] & $0.39 \pm 0.13$ & $64.28 \pm 1.84$ & $2.70 \pm 0.57$ & $52.74 \pm 1.76$ & $0.13$ & $58.94$ \\
Species error (detected) [\%] & $0.22 \pm 0.12$ & $63.96 \pm 1.81$ & $0.94 \pm 0.39$ & $51.99 \pm 1.62$ & $0.13$ & $52.68$ \\
Lattice length error [\AA] & $0.02 \pm 0.00$ & $0.08 \pm 0.04$ & $0.03 \pm 0.00$ & $0.06 \pm 0.01$ & $0.06$ & $0.10$ \\
Lattice angle error [deg] & $0.00 \pm 0.00$ & $0.00 \pm 0.00$ & $0.08 \pm 0.04$ & $0.35 \pm 0.02$ & $1.25$ & $0.05$ \\
\bottomrule
\end{tabular}
}
\end{table}

\begin{table}[btp]
\centering
\caption{\textbf{Reconstruction results (validation set).}}
\label{table:main_validation}
\scalebox{0.725}{
\begin{tabular}{lrrrrrr}
\toprule
Method & \multicolumn{1}{c}{Proposed} & \multicolumn{1}{c}{ICSG3D} & \multicolumn{1}{c}{Proposed} & \multicolumn{1}{c}{ICSG3D} & \multicolumn{1}{c}{Proposed} & \multicolumn{1}{c}{ICSG3D} \\
\midrule
Dataset & \multicolumn{2}{c}{ICSG3D} & \multicolumn{2}{c}{LCS6\AA} & \multicolumn{2}{c}{YBCO-like} \\
\midrule
Error in number of atoms [\%] & $0.85 \pm 0.31$ & $3.81 \pm 0.90$ & $6.33 \pm 0.63$ & $14.83 \pm 1.53$ & $14.00$ & $95.45$ \\
Position error (actual) [\AA] & $0.0334 \pm 0.0035$ & $0.1343 \pm 0.0319$ & $0.1503 \pm 0.0113$ & $0.2554 \pm 0.0276$ & $0.2935$ & $0.6752$ \\
Position error (detected) [\AA] & $0.0294 \pm 0.0034$ & $0.0985 \pm 0.0316$ & $0.1083 \pm 0.0083$ & $0.2015 \pm 0.0127$ & $0.2995$ & $0.4297$ \\
Species error (actual) [\%] & $3.75 \pm 0.35$ & $65.02 \pm 2.26$ & $13.22 \pm 0.81$ & $55.51 \pm 2.01$ & $21.15$ & $58.74$ \\
Species error (detected) [\%] & $3.65 \pm 0.36$ & $64.52 \pm 2.18$ & $12.22 \pm 0.81$ & $55.48 \pm 1.97$ & $20.72$ & $53.75$ \\
Lattice length error [\AA] & $0.02 \pm 0.00$ & $0.08 \pm 0.04$ & $0.05 \pm 0.00$ & $0.06 \pm 0.01$ & $0.34$ & $0.10$ \\
Lattice angle error [deg] & $0.00 \pm 0.00$ & $0.00 \pm 0.00$ & $0.15 \pm 0.03$ & $0.14 \pm 0.06$ & $3.75$ & $0.09$ \\
\bottomrule
\end{tabular}
}
\end{table}

As supplementary results for the quantitative performance evaluations in Section~2.3,
we show the reconstruction errors for the training and validation sets in Tables~\ref{table:main_train} and \ref{table:main_validation}, respectively.
The errors for the training set are lower than those for the test set, but the performance gaps between them seem reasonably small.

\section*{Supplementary Note 2. Additional results of latent space interpolation}
To supplement the interpolation analysis in Section~2.4, 
we here provide three additional examples.

\begin{figure*}[p]
\centering
\small
      \includegraphics[width=\textwidth]{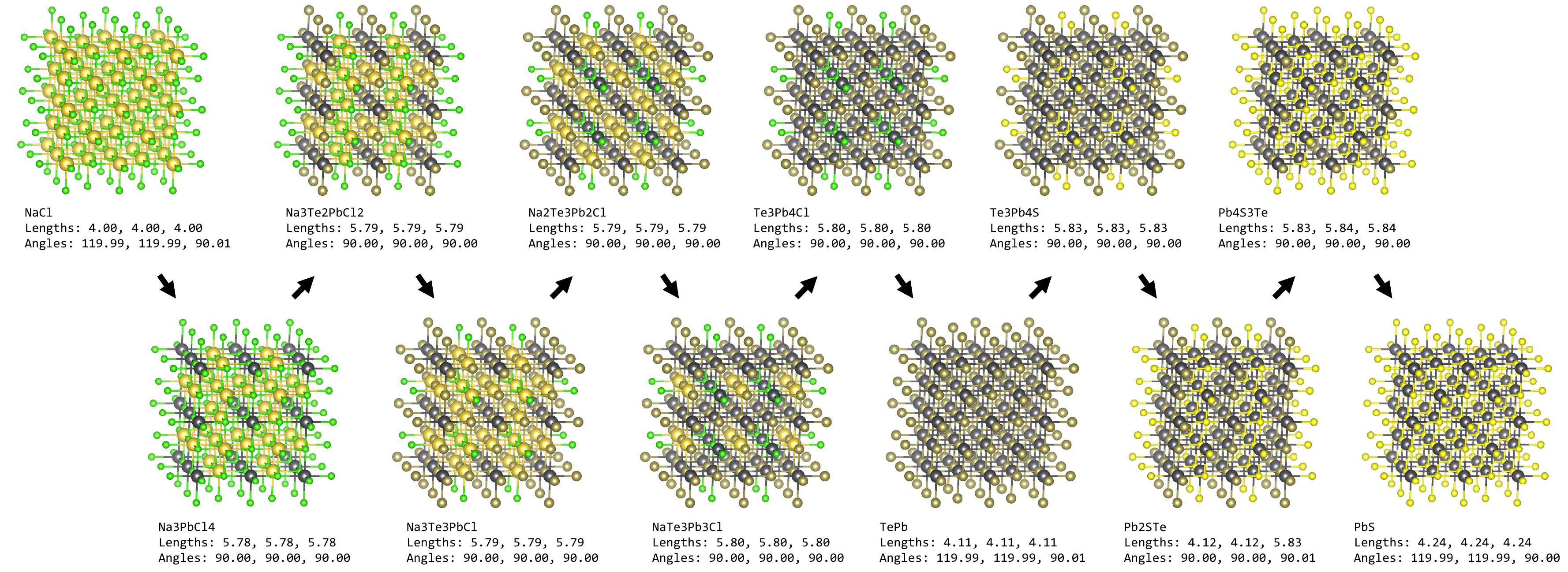}
      \caption{\textbf{Results of interpolation from \ce{NaCl} to \ce{PbS}.}}
      \label{fig:mophing_path3}
\end{figure*}

\begin{figure*}[p]
\centering
\small
      \includegraphics[width=0.7\textwidth]{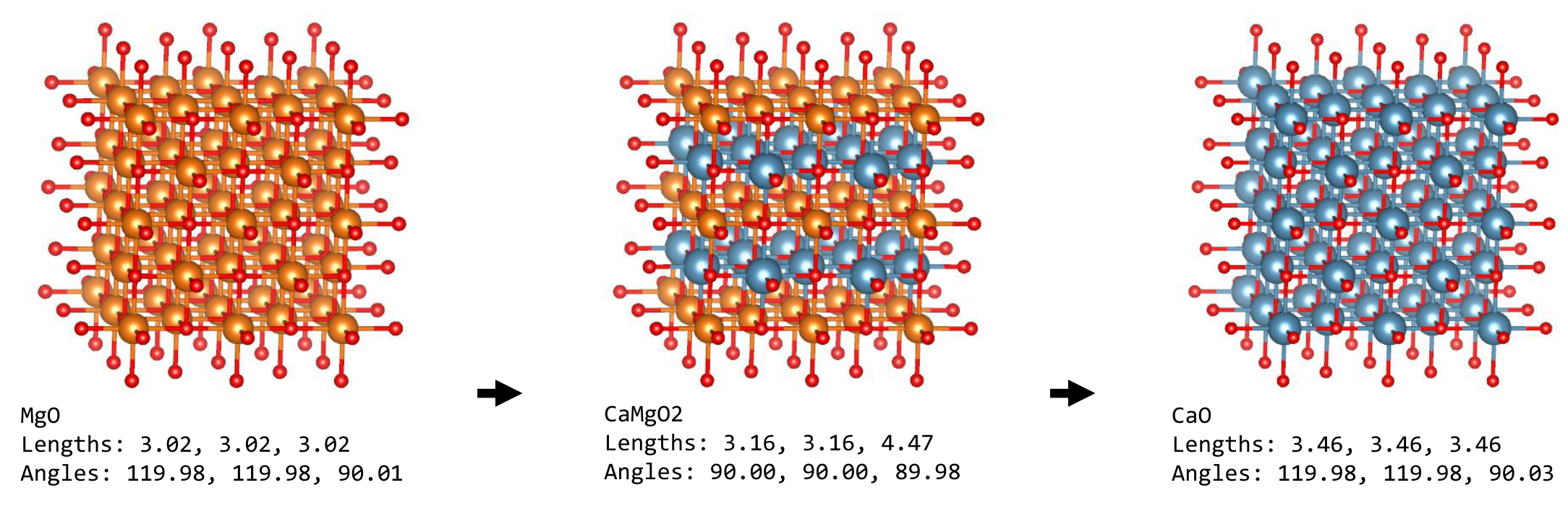}
      \caption{\textbf{Results of interpolation from \ce{MgO} to \ce{CaO}.}}
      \label{fig:mophing_path4}
\end{figure*}

\begin{figure*}[p]
\centering
\small
      \includegraphics[width=\textwidth]{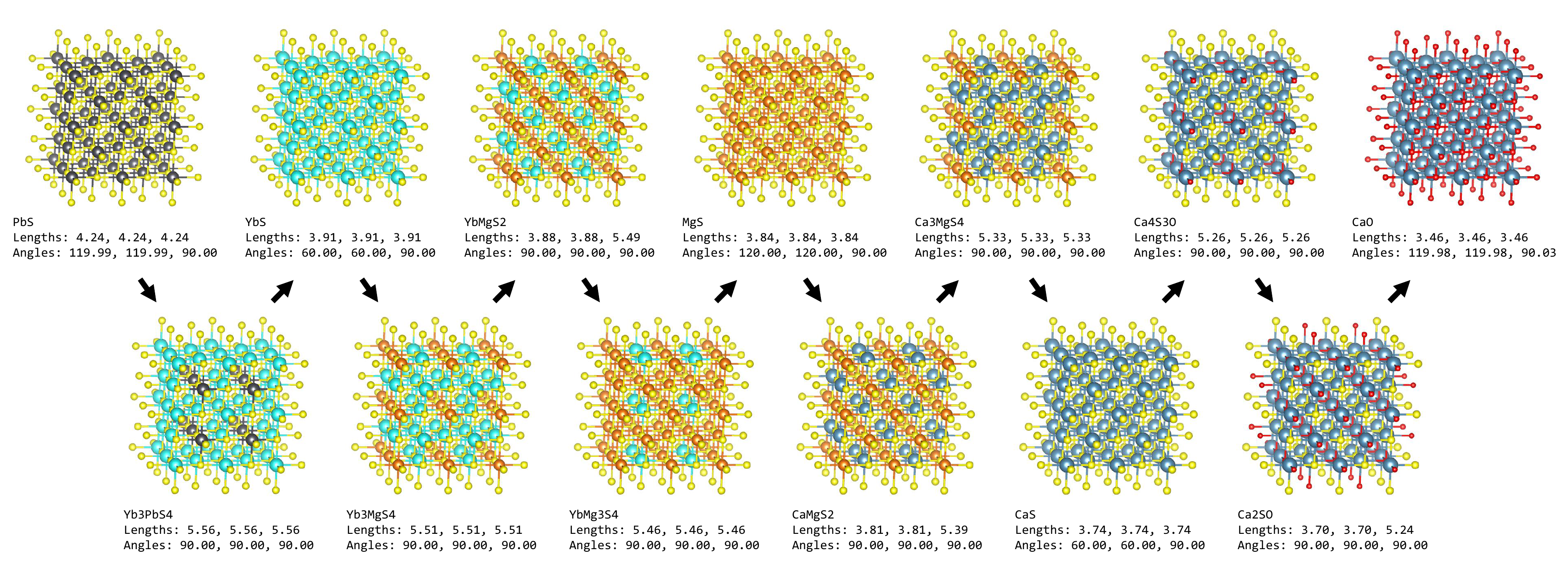}
      \caption{\textbf{Results of interpolation from \ce{PbS} to \ce{CaO}.}}
      \label{fig:mophing_path5}
\end{figure*}

\begin{figure*}[p]
\centering
\small
      \includegraphics[width=\textwidth]{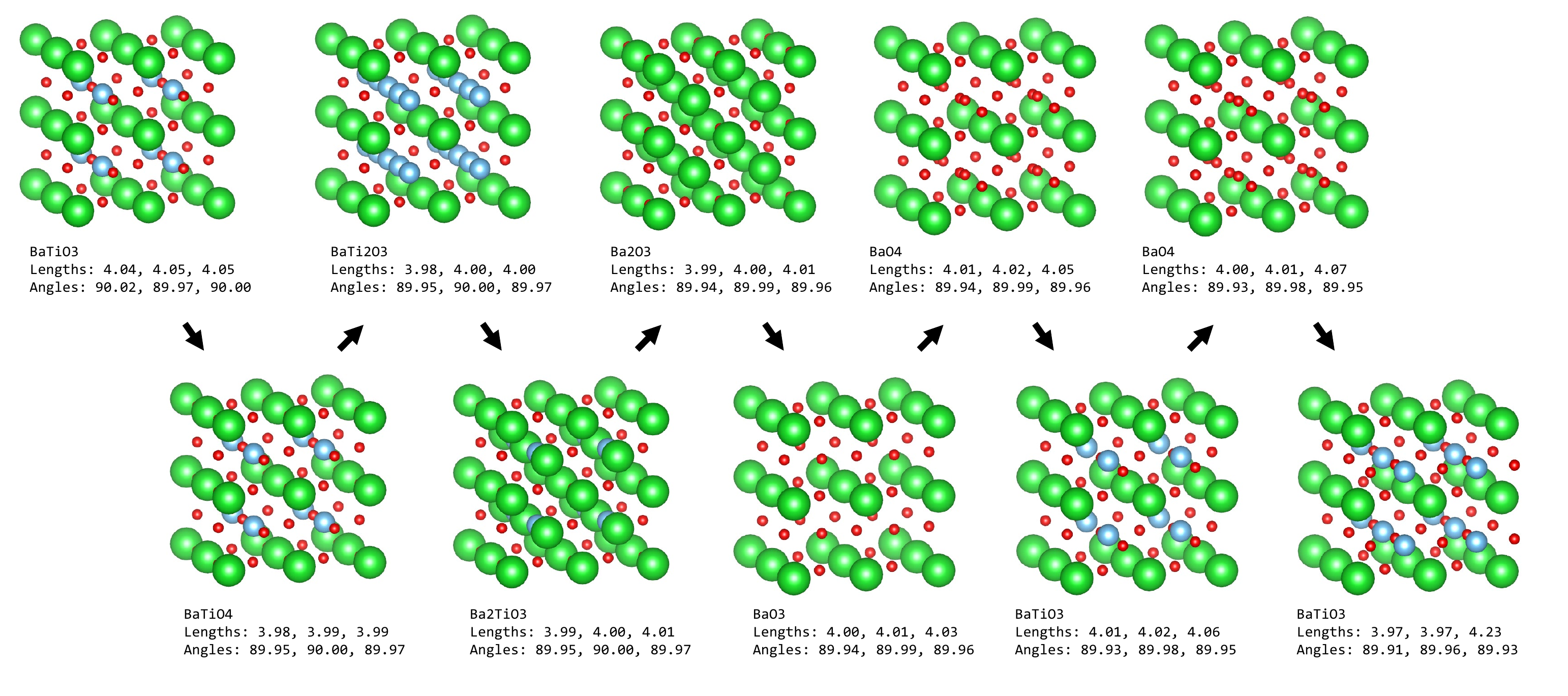}
      \caption{\textbf{Results of interpolation from \ce{BaTiO3} (cubic) to \ce{BaTiO3} (tetragonal).}}
      \label{fig:mophing_path6}
\end{figure*}

As a first additional example, Fig.~\ref{fig:mophing_path3} shows the results of interpolation from 
\ce{NaCl} (mp-22862) to \ce{PbS} (mp-21276). The obtained transition in the compositional formula is \ce{NaCl} $\to$ \ce{Na3PbCl4} (\ce{Na_{0.75}Pb_{0.25}Cl}) $\to$ \ce{Na3Te2PbCl2} (\ce{Na_{0.75}Pb_{0.25}Te_{0.5}Cl_{0.5}}) $\to$ \ce{Na3Te3PbCl} (\ce{Na_{0.75}Pb_{0.25}Pb_{0.75}Cl_{0.25}}) $\to$ \ce{Na2Te3Pb2Cl}\\ (\ce{Na_{0.5}Te_{0.5}Pb_{0.75}Cl_{0.25}}) $\to$ \ce{NaTe3Pb3Cl} (\ce{Na_{0.25}Te_{0.75}Pb_{0.75}Cl_{0.25}}) $\to$ \ce{Te3Pb4Cl} (\ce{PbTe_{0.75}Cl_{0.25}}) $\to$ \ce{TePb} $\to$ \ce{Te3Pb4S} (\ce{PbS_{0.25}Te_{0.75}}) $\to$ \ce{Pb2STe} (\ce{PbS_{0.5}Te_{0.5}}) $\to$ \ce{Pb4S3Te} (\ce{PbS_{0.75}Te_{0.25}}) $\to$ \ce{PbS}.

As a second additional example, Fig.~\ref{fig:mophing_path4} shows the results of inerpolation from \ce{MgO} (mp-1265) to \ce{CaO} (mp-2605). 
The obtained transition in the compositional formula is \ce{MgO} $\to$ \ce{CaMgO2} (\ce{Ca_{0.5}Mg_{0.5}O}) $\to$ \ce{CaO}.

As a third additional example, Fig.~\ref{fig:mophing_path5} shows the results of interpolation from \ce{PbS} (mp-21276) to \ce{CaO} (mp-2605). The obtained transition in the compositional formula is \ce{PbS} $\to$ \ce{Yb3PbS4}  (\ce{Yb_{0.75}Pb_{0.25}S}) $\to$ \ce{YbS} $\to$ \ce{Yb3MgS4} (\ce{Yb_{0.75}Mg_{0.25}S}) $\to$ \ce{YbMgS2} (\ce{Yb_{0.5}Mg_{0.5}S}) $\to$ \ce{YbMg3S4} (\ce{Yb_{0.25}Mg_{0.75}S}) $\to$ \ce{MgS} $\to$ \ce{CaMgS2} (\ce{Ca_{0.5}Mg_{0.5}S}) $\to$ \ce{Ca3MgS4} (\ce{Ca_{0.75}Mg_{0.25}S}) $\to$ \ce{CaS} $\to$ \ce{Ca4S3O} (\ce{CaS_{0.75}O_{0.25}}) $\to$ \ce{Ca2SO} (\ce{CaS_{0.5}O_{0.5}}) $\to$ \ce{CaO}.

As a fourth additional example, 
we chose to interpolate between \ce{BaTiO3} (cubic; mp-2998) and \ce{BaTiO3} (tetragonal; mp-5986) as a pair of structures with the same compositional formula but different Bravais lattices. 
Since the ICSG3D dataset consists only of cubic structures, we extracted their latent vectors from the LCS6{\AA} dataset using the model trained on it.
Fig.~\ref{fig:mophing_path6} shows the results of the interpolation.
The obtained transition in the compositional formula is \ce{BaTiO3} $\to$ \ce{BaTiO4} $\to$ \ce{BaTi2O3} $\to$ \ce{Ba2TiO3} $\to$ \ce{Ba2O3} $\to$ \ce{BaO3} $\to$ \ce{BaO4} $\to$ \ce{BaTiO3} $\to$ \ce{BaO4} $\to$ \ce{BaTiO3}. This result shows that the latent space can perform smooth transition even if the structures have different Bravais lattices.

\section*{Supplementary Note 3. Hyperparameters for crystal structure estimation}

\begin{figure*}[p]
\centering
\small
\includegraphics[height=0.9\textheight]{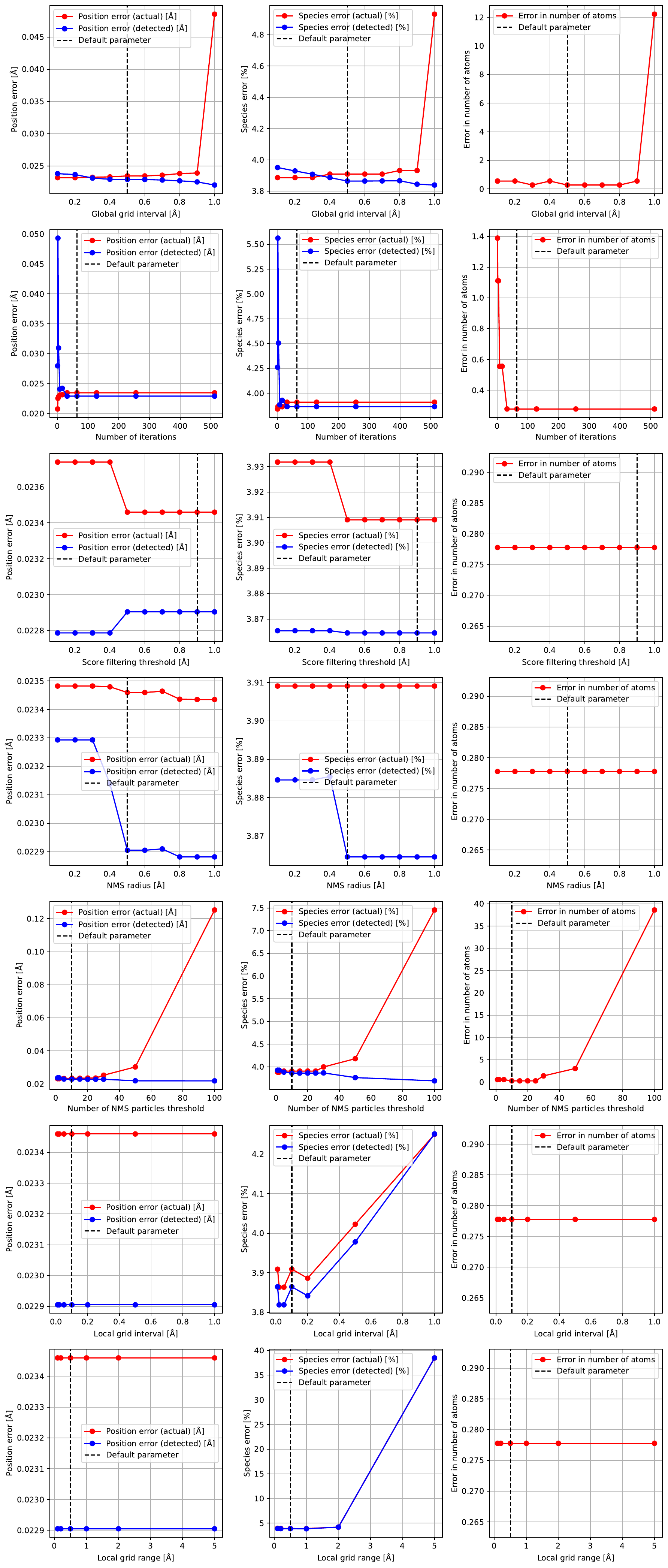}
\caption{\textbf{Hyperparameter sensitivity of the reconstruction algorithm for ICSG3D dataset.}  From top to bottom, we show the performance variations when changing one of the seven hyperparameters: the global grid interval, number of iterations, score filtering threshold, NMS radius, number of NMS particles threshold, local grid interval, and local grid range.
From left to right, we show the position error, species error, and error in the number of atoms. The default hyperparameter values are shown as vertical dashed lines.}
\label{fig:hyperparams_icsg3d}
\end{figure*}

\begin{figure*}[p]
\centering
\small
\includegraphics[height=0.9\textheight]{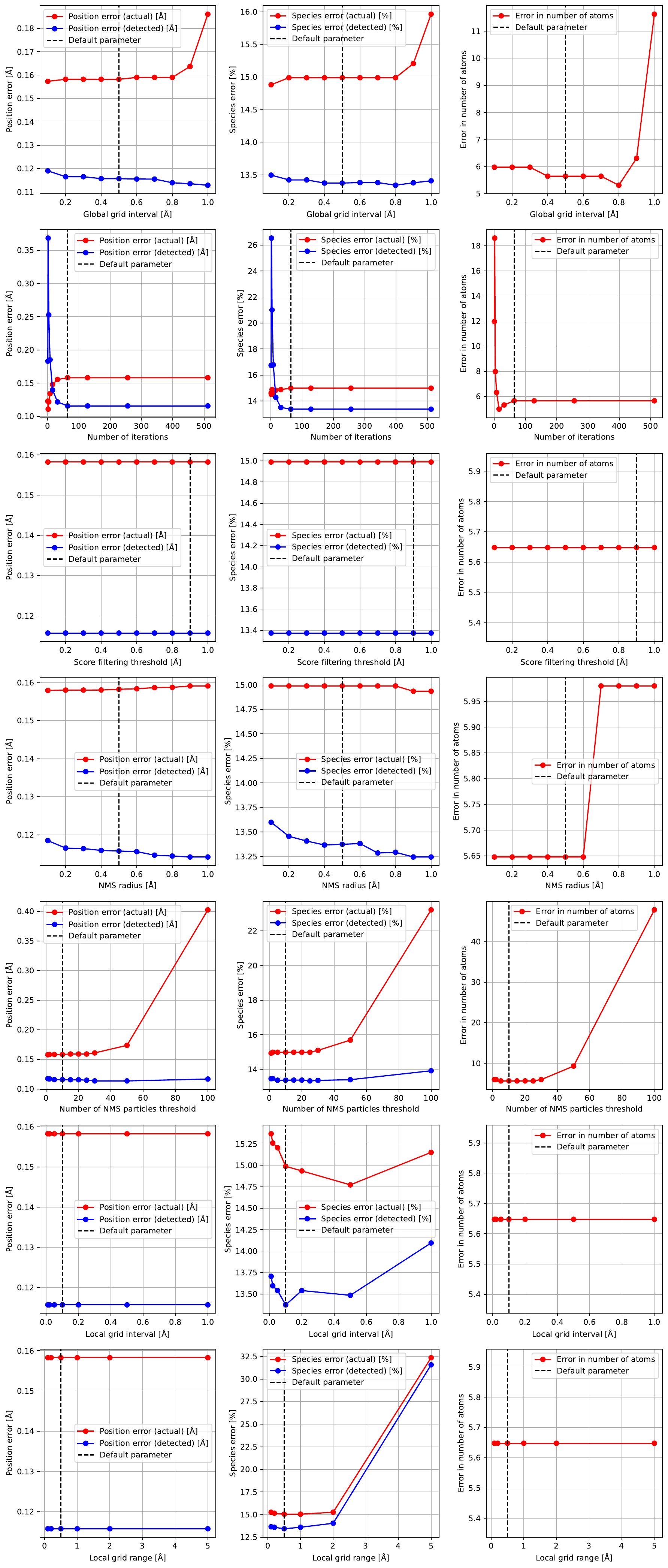}
\caption{\textbf{Hyperparameter sensitivity of the reconstruction algorithm for LCS6{\AA} dataset.}}
\label{fig:hyperparams_lcs6a}
\end{figure*}



The proposed reconstruction algorithm with NeSF has several hyperparameters, namely the global grid interval for particle initialization (step 1), the number of iterations of particle position update (step 2), the threshold for score filtering (step 3), the radius of the non-maximum suppression (NMS) (step 4), the threshold for the valid number of particles for the atom detection in the NMS (step 4), and the local grid interval and range for species estimation (step 5).

We have analyzed the sensitivity and validity of these hyperparameters.
Fig.~\ref{fig:hyperparams_icsg3d} and Fig.~\ref{fig:hyperparams_lcs6a} show the reconstruction errors for the ICSG3D and  LCS6{\AA} datasets, respectively, when changing one of these hyperparameters. The default hyperparameter values are shown as vertical dashed lines in the plots. These results confirm that the proposed algorithm is reasonably robust to these hyperparameter settings, and the chosen hyperparameters typically work well.


\section*{Supplementary Note 4. Prediction of material property}

\begin{table}[p]
\centering
\caption{\textbf{Result of material property prediction from the latent vectors.}}
\label{table:material_property_estimation}
\scalebox{0.6}[0.6]{
\begin{tabular}{@{}lllllll@{}}
\toprule
 & Property & Dataset & Split & Mean Square Error (MSE) & Mean Absolute Error (MAE) & R2 Score \\ \midrule
\multirow{18}{*}{Kernel SVR} & \multirow{6}{*}{Energy Above Hull} & \multirow{3}{*}{ICSG3D} & Train & 0.0558 & 0.1095 & 0.6003 \\ \cmidrule(l){4-7} 
 &  &  & Validation & 0.0441 & 0.1090 & 0.5386 \\ \cmidrule(l){4-7} 
 &  &  & Test & 0.0731 & 0.1303 & 0.5342 \\ \cmidrule(l){3-7} 
 &  & \multirow{3}{*}{LCS6{\AA}} & Train & 6.1224 & 0.8989 & -0.0015 \\ \cmidrule(l){4-7} 
 &  &  & Validation & 6.4173 & 0.9469 & -0.0175 \\ \cmidrule(l){4-7} 
 &  &  & Test & 5.8475 & 0.9085 & -0.0273 \\ \cmidrule(l){2-7} 
 & \multirow{6}{*}{Total Magnetization} & \multirow{3}{*}{ICSG3D} & Train & 0.3860 & 0.1627 & 0.2194 \\ \cmidrule(l){4-7} 
 &  &  & Validation & 0.5227 & 0.1939 & 0.1287 \\ \cmidrule(l){4-7} 
 &  &  & Test & 0.2498 & 0.1474 & 0.2020 \\ \cmidrule(l){3-7} 
 &  & \multirow{3}{*}{LCS6{\AA}} & Train & 0.0886 & 0.1535 & 0.5425 \\ \cmidrule(l){4-7} 
 &  &  & Validation & 0.1297 & 0.1991 & 0.3553 \\ \cmidrule(l){4-7} 
 &  &  & Test & 0.0870 & 0.1641 & 0.3435 \\ \cmidrule(l){2-7} 
 & \multirow{6}{*}{Bandgap} & \multirow{3}{*}{ICSG3D} & Train & 7.3111 & 1.0413 & 0.0157 \\ \cmidrule(l){4-7} 
 &  &  & Validation & 8.7110 & 1.2267 & -0.0042 \\ \cmidrule(l){4-7} 
 &  &  & Test & 8.0715 & 1.0829 & 0.0415 \\ \cmidrule(l){3-7} 
 &  & \multirow{3}{*}{LCS6{\AA}} & Train & 1.0300 & 0.3572 & 0.2735 \\ \cmidrule(l){4-7} 
 &  &  & Validation & 0.7289 & 0.3440 & 0.2259 \\ \cmidrule(l){4-7} 
 &  &  & Test & 0.8744 & 0.3569 & 0.1520 \\ \midrule
\multirow{18}{*}{kNN} & \multirow{6}{*}{Energy Above Hull} & \multirow{3}{*}{ICSG3D} & Train & 0.0000 & 0.0000 & 1.0000 \\ \cmidrule(l){4-7} 
 &  &  & Validation & 0.0490 & 0.0893 & 0.4873 \\ \cmidrule(l){4-7} 
 &  &  & Test & 0.0733 & 0.1027 & 0.5324 \\ \cmidrule(l){3-7} 
 &  & \multirow{3}{*}{LCS6{\AA}} & Train & 0.0000 & 0.0000 & 1.0000 \\ \cmidrule(l){4-7} 
 &  &  & Validation & 2.3434 & 0.7649 & 0.6285 \\ \cmidrule(l){4-7} 
 &  &  & Test & 2.5413 & 0.7936 & 0.5536 \\ \cmidrule(l){2-7} 
 & \multirow{6}{*}{Total Magnetization} & \multirow{3}{*}{ICSG3D} & Train & 0.0000 & 0.0000 & 1.0000 \\ \cmidrule(l){4-7} 
 &  &  & Validation & 0.2788 & 0.1521 & 0.5352 \\ \cmidrule(l){4-7} 
 &  &  & Test & 0.1897 & 0.1225 & 0.3937 \\ \cmidrule(l){3-7} 
 &  & \multirow{3}{*}{LCS6{\AA}} & Train & 0.0000 & 0.0000 & 1.0000 \\ \cmidrule(l){4-7} 
 &  &  & Validation & 0.1345 & 0.1848 & 0.3310 \\ \cmidrule(l){4-7} 
 &  &  & Test & 0.0752 & 0.1368 & 0.4325 \\ \cmidrule(l){2-7} 
 & \multirow{6}{*}{Bandgap} & \multirow{3}{*}{ICSG3D} & Train & 0.0000 & 0.0000 & 1.0000 \\ \cmidrule(l){4-7} 
 &  &  & Validation & 3.5833 & 1.1590 & 0.5869 \\ \cmidrule(l){4-7} 
 &  &  & Test & 3.2746 & 1.0944 & 0.6111 \\ \cmidrule(l){3-7} 
 &  & \multirow{3}{*}{LCS6{\AA}} & Train & 0.0000 & 0.0000 & 1.0000 \\ \cmidrule(l){4-7} 
 &  &  & Validation & 0.8403 & 0.3878 & 0.1076 \\ \cmidrule(l){4-7} 
 &  &  & Test & 0.7947 & 0.3800 & 0.2293 \\ \midrule
\multirow{18}{*}{MLP} & \multirow{6}{*}{Energy Above Hull} & \multirow{3}{*}{ICSG3D} & Train & 0.0035 & 0.0397 & 0.9751 \\ \cmidrule(l){4-7} 
 &  &  & Validation & 0.0266 & 0.0866 & 0.7217 \\ \cmidrule(l){4-7} 
 &  &  & Test & 0.0549 & 0.1085 & 0.6497 \\ \cmidrule(l){3-7} 
 &  & \multirow{3}{*}{LCS6{\AA}} & Train & 0.0950 & 0.1990 & 0.9845 \\ \cmidrule(l){4-7} 
 &  &  & Validation & 1.1107 & 0.5404 & 0.8239 \\ \cmidrule(l){4-7} 
 &  &  & Test & 1.6265 & 0.6113 & 0.7143 \\ \cmidrule(l){2-7} 
 & \multirow{6}{*}{Total Magnetization} & \multirow{3}{*}{ICSG3D} & Train & 0.0323 & 0.0767 & 0.9348 \\ \cmidrule(l){4-7} 
 &  &  & Validation & 0.2723 & 0.1786 & 0.5462 \\ \cmidrule(l){4-7} 
 &  &  & Test & 0.1151 & 0.1290 & 0.6323 \\ \cmidrule(l){3-7} 
 &  & \multirow{3}{*}{LCS6{\AA}} & Train & 0.0030 & 0.0394 & 0.9847 \\ \cmidrule(l){4-7} 
 &  &  & Validation & 0.1155 & 0.1860 & 0.4254 \\ \cmidrule(l){4-7} 
 &  &  & Test & 0.0739 & 0.1557 & 0.4428 \\ \cmidrule(l){2-7} 
 & \multirow{6}{*}{Bandgap} & \multirow{3}{*}{ICSG3D} & Train & 0.5132 & 0.4314 & 0.9309 \\ \cmidrule(l){4-7} 
 &  &  & Validation & 3.6280 & 1.0056 & 0.5818 \\ \cmidrule(l){4-7} 
 &  &  & Test & 3.3493 & 0.9423 & 0.6022 \\ \cmidrule(l){3-7} 
 &  & \multirow{3}{*}{LCS6{\AA}} & Train & 0.0399 & 0.1114 & 0.9719 \\ \cmidrule(l){4-7} 
 &  &  & Validation & 0.6426 & 0.3553 & 0.3175 \\ \cmidrule(l){4-7} 
 &  &  & Test & 0.6577 & 0.3853 & 0.3622 \\ \bottomrule
\end{tabular}
}
\end{table}

In this analysis, we investigated whether the latent space captures the physical properties of the materials even when it is not trained to do so.
Given the pre-trained latent vectors as input, we used three basic regression methods, Kernel SVR, kNN, and MLP, to predict material properties.
For Kernel SVR, we used the radial basis kernel function, and set the regularization parameter $C$ to $1.0$ and $\epsilon$ to $0.1$. For kNN, we used five nearest neighbors ($k = 5$). For MLP, we used four fully-connected layers with the ReLU activation function and the dimensions of the layers were 192 (input) 128, 128, 64, and 1 (final output). We trained the MLP using the Adam optimizer with the constant learning rate of $0.001$ and the batch size of 200 for 1000 epochs. Most of these hyperparameters follow the default settings of the scikit-learn implementation.


The property prediction results are summarized in Table~\ref{table:material_property_estimation}.
The results show that, although we observe relatively high test R2 scores (higher than 0.5) in some cases, the latent vectors are not generally suitable for property prediction. This result suggests that crystal structures and their physical properties are entangled via complicated relationships, which could not be easily disentangled in the task of auto-encoding crystal structures alone. Making the latent space more effectively disentangle physical properties from structural features is part of future work.

\section*{Supplementary Note 5. Network architecture}
The proposed network architecture is specified in the following list, where $N$ is the number of atoms in the input crystal structure, $A$ is the number of atomic species in the dataset, FC stands for fully connected layers, and numbers in layers represent their feature dimensions.
\begin{itemize}
  \item Encoder
  \begin{itemize}
    \item Atom position encoder
    \begin{itemize}
      \item Input: Atom positions: [Nx3]
      \item Shared MLP:  [Nx3] -> [Nx128]
      \begin{itemize}
        \item Layers: 3, 64, 128, 128
        \item Activation function: ReLU
        \item Using batch normalization
        \item No dropout
      \end{itemize}
      \item Output: Atom position features: [Nx128]
    \end{itemize}
    \item Atom species encoder
    \begin{itemize}
      \item Input: Atom species with one-hot encodings: [NxA]
      \item Shared MLP: [NxA] -> [Nx128]
      \begin{itemize}
        \item Layers: A, 128, 128
        \item Activation function: ReLU
        \item Using batch normalization
        \item No dropout
      \end{itemize}
      \item Output: Atom species features: [Nx128]
    \end{itemize}
    \item Crystal structure encoder
    \begin{itemize}
      \item Input: Atom position features and atom species features: [Nx128], [Nx128]
      \item Concatenation of atom position features and atom species features: [Nx128], [Nx128] -> [Nx256]
      \item Shared MLP: [Nx256] -> [Nx512]
      \begin{itemize}
        \item Layers: 256, 512
        \item Activation function: ReLU
        \item Using dropout with 0.3 dropout probability
      \end{itemize}
      \item Max-Pooling (along the atoms): [Nx512] -> [512]
      \item Concatenation with lattice parameters (length and angles): [512], [6] -> [518]
      \item MLP: [518] -> [192]
      \begin{itemize}
        \item Layers: 518, 256, 256, 192
        \item Activation function: ReLU
        \item Using batch normalization
        \item Using dropout with 0.3 dropout probability
      \end{itemize}
      \item Output: Latent vector: [192]
    \end{itemize}
  \end{itemize}
  \item Decoder
  \begin{itemize}
    \item Lattice lenght parameter decoder
    \begin{itemize}
      \item Input: Latent vector: [192]
      \item MLP: [192] -> [3]
      \begin{itemize}
        \item Layers: 192, 128, 3
        \item Activation function: ReLU
        \item No batch normalization
        \item No dropout
      \end{itemize}
      \item Output: Lattice length parameter: [3]
    \end{itemize}
    \item Lattice angle parameter decoder
    \begin{itemize}
      \item Input: Latent vector: [192]
      \item MLP: [192] -> [3]
      \begin{itemize}
        \item Layers: 192, 128, 3
        \item Activation function: ReLU
        \item No batch normalization
        \item No dropout
      \end{itemize}
      \item Output: Lattice angle parameter: [3]
    \end{itemize}
    \item Position field decoder:
    \begin{itemize}
      \item Input: Latent vector and query position: [192], [3]
      \item Concatenation of Latent vector and query position: [192], [3] -> [195]
      \item FC: [195] -> [256]
      \begin{itemize}
        \item Activation function: ReLU
        \item Using batch normalization
        \item No dropout
      \end{itemize}
      \item Concatenation of above output and query position: [256], [3] -> [259]
      \item FC: [259] -> [128]
      \begin{itemize}
        \item Activation function: ReLU
        \item Using batch normalization
        \item No dropout
      \end{itemize}
      \item Concatenation of above output and query position: [128], [3] -> [131]
      \item FC: [131] -> [128]
      \begin{itemize}
        \item Activation function: ReLU
        \item Using batch normalization
        \item No dropout
      \end{itemize}
      \item Concatenation of above output and query position: [128], [3] -> [131]
      \item FC: [131] -> [128]
      \begin{itemize}
        \item Activation function: ReLU
        \item Using batch normalization
        \item No dropout
      \end{itemize}
      \item Concatenation of above output and query position: [128], [3] -> [131]
      \item FC: [131] -> [128]
      \begin{itemize}
        \item Activation function: ReLU
        \item Using batch normalization
        \item No dropout
      \end{itemize}
      \item Concatenation of above output and query position: [128], [3] -> [131]
      \item FC: [131] -> [128]
      \begin{itemize}
        \item Activation function: ReLU
        \item Using batch normalization
        \item No dropout
      \end{itemize}
      \item Concatenation of above output and query position: [128], [3] -> [131]
      \item FC: [131] -> [128]
      \begin{itemize}
        \item Activation function: ReLU
        \item Using batch normalization
        \item No dropout
      \end{itemize}
      \item Concatenation of above output and query position: [128], [3] -> [131]
      \item FC: [131] -> [128]
      \begin{itemize}
        \item Activation function: ReLU
        \item Using batch normalization
        \item No dropout
      \end{itemize}
      \item Concatenation of above output and query position: [128], [3] -> [131]
      \item FC: [131] -> [3]
      \begin{itemize}
        \item Activation function: None
        \item No batch normalization
        \item No dropout
      \end{itemize}
      \item Output: Position field value of the query position: [3]
    \end{itemize}
    \item Species field decoder:
    \begin{itemize}
      \item Input: Latent vector and query position: [192], [3]
      \item Concatenation of Latent vector and query position: [192], [3] -> [195]
      \item FC: [195] -> [256]
      \begin{itemize}
        \item Activation function: ReLU
        \item Using batch normalization
        \item No dropout
      \end{itemize}
      \item Concatenation of above output and query position: [256], [3] -> [259]
      \item FC: [259] -> [128]
      \begin{itemize}
        \item Activation function: ReLU
        \item Using batch normalization
        \item No dropout
      \end{itemize}
      \item Concatenation of above output and query position: [128], [3] -> [131]
      \item FC: [131] -> [A]
      \begin{itemize}
        \item Activation function: None
        \item No batch normalization
        \item No dropout
      \end{itemize}
      \item Output: Species field value (class likelihood) of the query position: [A]
    \end{itemize}
  \end{itemize}
\end{itemize}

\section*{Supplementary Note 6. Sampling parameters}
For the training of the NeSF and crystal structure estimation, we use the global and local grid sampling with the following parameters.

\begin{itemize}
\item For Training:
\begin{itemize}
  \item For the position field:
  \begin{itemize}
    \item Global grid sampling:
    \begin{itemize}
      \item Grid interval: $1$ \AA
      \item Perturbation scale $\sigma$: $0.5$ \AA
    \end{itemize}
    \item Local grid sampling:
    \begin{itemize}
      \item Grid interval: $0.3$ \AA
      \item Grid range for each axis: $1.2$ \AA
      \item Perturbation scale $\sigma$: $0.5$ \AA
    \end{itemize}
  \end{itemize}
  \item For the species field:
  \begin{itemize}
    \item Local grid sampling:
    \begin{itemize}
      \item Grid interval: $0.1$ \AA
      \item Grid range for each axis: $0.5$ \AA
      \item Perturbation scale $\sigma$: $0.5$ \AA
    \end{itemize}
  \end{itemize}
\end{itemize}
\item For crystal structure estimation:
\begin{itemize}
  \item For the position field:
  \begin{itemize}
    \item Global grid sampling:
    \begin{itemize}
      \item Grid interval: $0.5$ \AA
    \end{itemize}
  \end{itemize}
  \item For the species field:
  \begin{itemize}
    \item Local grid sampling:
    \begin{itemize}
      \item Grid interval: $0.1$ \AA
      \item Grid range for each axis: $0.5$ \AA
    \end{itemize}
  \end{itemize}
\end{itemize}
\end{itemize}

\section*{Supplementary Note 7. Hyperparameter search}
The hyperparameter search procedure is summarized as the eight steps below. In each step, we evaluated multiple configurations of certain hyperparameters by training and validating on the LCS6{\AA} dataset and chose the best configuration for the next step. 
We evaluated the error in number of atoms as the primal performance indicator. When the performance was the same by this metric, the position error (actual) was used as the secondary performance indicator. The final step (step 8) is to retune some hyperparameters but it could not improve the performance. 

\begin{enumerate}
\scriptsize
  \item[0.] Target parameters and the initial configuration
  \begin{itemize}
      \item \text{Weights of loss functions: loss\_weight.pos=1.0, loss\_weight.spec=1.0, loss\_weight.abc=1.0, loss\_weight.angle=1.0}
      \item \text{The dimension of the latent vector: latent\_size=256}
      \item \text{The learning rate of the Adam optimizer: training.lr=0.001}
      \item Layers of the position field decoder: \text{decoder.pos.channels=[256,128,128,128,128,128,128,128,128]}
      \item Layers of the species field decoder: \text{decoder.spec.channels=[256,128,128,128,128,128,128,128,128]}
      \item Layers of the atom position encoder: \text{encoder.smlp\_pos.channels=[64,128]}
      \item Layers of the atom species encoder: \text{encoder.smlp\_spec.channels=[128]}
      \item Layers of the crystal structure encoder: \text{encoder.smlp\_mix.channels=[512], encoder.mlp.channels=[256]}
      \item Batch size: \text{batch\_size=64}
      
  \end{itemize}
  \item Weights of loss functions
  \begin{itemize}
    \item \text{loss\_weight.pos=0.1, loss\_weight.spec=0.1, loss\_weight.abc=1.0, loss\_weight.angle=1.0}
    \item \text{loss\_weight.pos=0.1, loss\_weight.spec=1.0, loss\_weight.abc=1.0, loss\_weight.angle=1.0}
    \item \text{loss\_weight.pos=0.1, loss\_weight.spec=10.0, loss\_weight.abc=1.0, loss\_weight.angle=1.0}
    \item \text{loss\_weight.pos=1.0, loss\_weight.spec=0.1, loss\_weight.abc=1.0, loss\_weight.angle=1.0}
    \item \text{loss\_weight.pos=1.0, loss\_weight.spec=1.0, loss\_weight.abc=1.0, loss\_weight.angle=1.0}
    \item \text{loss\_weight.pos=1.0, loss\_weight.spec=10.0, loss\_weight.abc=1.0, loss\_weight.angle=1.0}
    \item \text{loss\_weight.pos=10.0, loss\_weight.spec=0.1, loss\_weight.abc=1.0, loss\_weight.angle=1.0}
    \item \text{loss\_weight.pos=10.0, loss\_weight.spec=1.0, loss\_weight.abc=1.0, loss\_weight.angle=1.0}
    \item \text{loss\_weight.pos=10.0, loss\_weight.spec=10.0, loss\_weight.abc=1.0, loss\_weight.angle=1.0}
  \end{itemize}
  \item Size of latent vector
  \begin{itemize}
    \item \text{latent\_size=64}
    \item \text{latent\_size=92}
    \item \text{latent\_size=128}
    \item \text{latent\_size=192}
    \item \text{latent\_size=256}
    \item \text{latent\_size=384}
    \item \text{latent\_size=512}
    \item \text{latent\_size=1024}
  \end{itemize}
  \item Learning rate
  \begin{itemize}
    \item \text{training.lr=1.0}
    \item \text{training.lr=0.1}
    \item \text{training.lr=0.01}
    \item \text{training.lr=0.005}
    \item \text{training.lr=0.001}
    \item \text{training.lr=0.0005}
    \item \text{training.lr=0.0001}
    \item \text{training.lr=0.00001}
  \end{itemize}
  \item Decoder MLP
  \begin{itemize}
    \item \text{decoder.pos.channels=[256,128,128,128,128,128,128,128], decoder.spec.channels=[256,128,128,128,128,128,128,128]}
    \item \text{decoder.pos.channels=[256,128,128,128,128,128,128,128], decoder.spec.channels=[256,128,128,128,128,128]}
    \item \text{decoder.pos.channels=[256,128,128,128,128,128,128,128], decoder.spec.channels=[256,128,128,128,128]}
    \item \text{decoder.pos.channels=[256,128,128,128,128,128,128,128], decoder.spec.channels=[256,128,128,128]}
    \item \text{decoder.pos.channels=[256,128,128,128,128,128,128,128], decoder.spec.channels=[256,128]}
    \item \text{decoder.pos.channels=[256,128,128,128,128,128], decoder.spec.channels=[256,128,128,128,128,128,128,128]}
    \item \text{decoder.pos.channels=[256,128,128,128,128,128], decoder.spec.channels=[256,128,128,128,128,128]}
    \item \text{decoder.pos.channels=[256,128,128,128,128,128], decoder.spec.channels=[256,128,128,128,128]}
    \item \text{decoder.pos.channels=[256,128,128,128,128,128], decoder.spec.channels=[256,128,128,128]}
    \item \text{decoder.pos.channels=[256,128,128,128,128,128], decoder.spec.channels=[256,128]}
    \item \text{decoder.pos.channels=[256,128,128,128,128], decoder.spec.channels=[256,128,128,128,128,128,128,128]}
    \item \text{decoder.pos.channels=[256,128,128,128,128], decoder.spec.channels=[256,128,128,128,128,128]}
    \item \text{decoder.pos.channels=[256,128,128,128,128], decoder.spec.channels=[256,128,128,128,128]}
    \item \text{decoder.pos.channels=[256,128,128,128,128], decoder.spec.channels=[256,128,128,128]}
    \item \text{decoder.pos.channels=[256,128,128,128,128], decoder.spec.channels=[256,128]}
    \item \text{decoder.pos.channels=[256,128,128,128], decoder.spec.channels=[256,128,128,128,128,128,128,128]}
    \item \text{decoder.pos.channels=[256,128,128,128], decoder.spec.channels=[256,128,128,128,128,128]}
    \item \text{decoder.pos.channels=[256,128,128,128], decoder.spec.channels=[256,128,128,128,128]}
    \item \text{decoder.pos.channels=[256,128,128,128], decoder.spec.channels=[256,128,128,128]}
    \item \text{decoder.pos.channels=[256,128], decoder.spec.channels=[256,128,128,128,128,128,128,128]}
    \item \text{decoder.pos.channels=[256,128], decoder.spec.channels=[256,128,128,128,128,128]}
    \item \text{decoder.pos.channels=[256,128], decoder.spec.channels=[256,128,128,128,128]}
    \item \text{decoder.pos.channels=[256,128,128,128], decoder.spec.channels=[256,128]}
    \item \text{decoder.pos.channels=[256,128], decoder.spec.channels=[256,128,128,128]}
    \item \text{decoder.pos.channels=[256,128], decoder.spec.channels=[256,128]}
  \end{itemize}
  \item Encoder MLP
  \begin{itemize}
    \item \text{encoder.smlp\_pos.channels=[64,128], encoder.smlp\_spec.channels=[128]}
    \item \text{encoder.smlp\_pos.channels=[64,128], encoder.smlp\_spec.channels=[128,128,128]}
    \item \text{encoder.smlp\_pos.channels=[64,128,128], encoder.smlp\_spec.channels=[128]}
    \item \text{encoder.smlp\_pos.channels=[64,128,128], encoder.smlp\_spec.channels=[128,128]}
    \item \text{encoder.smlp\_pos.channels=[64,128,128], encoder.smlp\_spec.channels=[128,128,128]}
    \item \text{encoder.smlp\_pos.channels=[64,128,128,128], encoder.smlp\_spec.channels=[128]}
    \item \text{encoder.smlp\_pos.channels=[64,128,128,128], encoder.smlp\_spec.channels=[128,128]}
    \item \text{encoder.smlp\_pos.channels=[64,128,128,128], encoder.smlp\_spec.channels=[128,128,128]}
    \item \text{encoder.smlp\_pos.channels=[64,128], encoder.smlp\_spec.channels=[128,128]}
  \end{itemize}
  \item Crystal structure encoder
  \begin{itemize}
    \item \text{encoder.smlp\_mix.channels=[256], encoder.mlp.channels=[512]}
    \item \text{encoder.smlp\_mix.channels=[256], encoder.mlp.channels=[512,512,256]}
    \item \text{encoder.smlp\_mix.channels=[256,256], encoder.mlp.channels=[512]}
    \item \text{encoder.smlp\_mix.channels=[256,256], encoder.mlp.channels=[512,512]}
    \item \text{encoder.smlp\_mix.channels=[256,256], encoder.mlp.channels=[512,512,256]}
    \item \text{encoder.smlp\_mix.channels=[256,256,512], encoder.mlp.channels=[512]}
    \item \text{encoder.smlp\_mix.channels=[256,256,512], encoder.mlp.channels=[512,512]}
    \item \text{encoder.smlp\_mix.channels=[256,256,512], encoder.mlp.channels=[512,512,256]}
    \item \text{encoder.smlp\_mix.channels=[256], encoder.mlp.channels=[512,512]}
  \end{itemize}
  \item Batch size
  \begin{itemize}
    \item \text{batch\_size=2}
    \item \text{batch\_size=4}
    \item \text{batch\_size=8}
    \item \text{batch\_size=16}
    \item \text{batch\_size=32}
    \item \text{batch\_size=64}
    \item \text{batch\_size=128}
    \item \text{batch\_size=256}
  \end{itemize}
  \item Total verification
  \begin{itemize}
    \item \text{latent\_size=192, training.lr=0.01, loss\_weight.pos=0.1, loss\_weight.spec=0.1}
    \item \text{latent\_size=192, training.lr=0.01, loss\_weight.pos=0.1, loss\_weight.spec=1.0}
    \item \text{latent\_size=192, training.lr=0.01, loss\_weight.pos=0.1, loss\_weight.spec=10.0}
    \item \text{latent\_size=192, training.lr=0.01, loss\_weight.pos=1.0, loss\_weight.spec=0.1}
    \item \text{latent\_size=192, training.lr=0.01, loss\_weight.pos=1.0, loss\_weight.spec=1.0}
    \item \text{latent\_size=192, training.lr=0.01, loss\_weight.pos=1.0, loss\_weight.spec=10.0}
    \item \text{latent\_size=192, training.lr=0.01, loss\_weight.pos=10.0, loss\_weight.spec=0.1}
    \item \text{latent\_size=192, training.lr=0.01, loss\_weight.pos=10.0, loss\_weight.spec=1.0}
    \item \text{latent\_size=192, training.lr=0.01, loss\_weight.pos=10.0, loss\_weight.spec=10.0}
    \item \text{latent\_size=192, training.lr=0.001, loss\_weight.pos=0.1, loss\_weight.spec=0.1}
    \item \text{latent\_size=192, training.lr=0.001, loss\_weight.pos=0.1, loss\_weight.spec=1.0}
    \item \text{latent\_size=192, training.lr=0.001, loss\_weight.pos=0.1, loss\_weight.spec=10.0}
    \item \text{latent\_size=192, training.lr=0.001, loss\_weight.pos=1.0, loss\_weight.spec=0.1}
    \item \text{latent\_size=192, training.lr=0.001, loss\_weight.pos=1.0, loss\_weight.spec=1.0}
    \item \text{latent\_size=192, training.lr=0.001, loss\_weight.pos=1.0, loss\_weight.spec=10.0}
    \item \text{latent\_size=192, training.lr=0.001, loss\_weight.pos=10.0, loss\_weight.spec=0.1}
    \item \text{latent\_size=192, training.lr=0.001, loss\_weight.pos=10.0, loss\_weight.spec=1.0}
    \item \text{latent\_size=192, training.lr=0.001, loss\_weight.pos=10.0, loss\_weight.spec=10.0}
    \item \text{latent\_size=192, training.lr=0.0001, loss\_weight.pos=0.1, loss\_weight.spec=0.1}
    \item \text{latent\_size=192, training.lr=0.0001, loss\_weight.pos=0.1, loss\_weight.spec=1.0}
    \item \text{latent\_size=192, training.lr=0.0001, loss\_weight.pos=0.1, loss\_weight.spec=10.0}
    \item \text{latent\_size=192, training.lr=0.0001, loss\_weight.pos=1.0, loss\_weight.spec=0.1}
    \item \text{latent\_size=192, training.lr=0.0001, loss\_weight.pos=1.0, loss\_weight.spec=1.0}
    \item \text{latent\_size=192, training.lr=0.0001, loss\_weight.pos=1.0, loss\_weight.spec=10.0}
    \item \text{latent\_size=192, training.lr=0.0001, loss\_weight.pos=10.0, loss\_weight.spec=0.1}
    \item \text{latent\_size=192, training.lr=0.0001, loss\_weight.pos=10.0, loss\_weight.spec=1.0}
    \item \text{latent\_size=192, training.lr=0.0001, loss\_weight.pos=10.0, loss\_weight.spec=10.0}
    \item \text{latent\_size=256, training.lr=0.01, loss\_weight.pos=0.1, loss\_weight.spec=0.1}
    \item \text{latent\_size=256, training.lr=0.01, loss\_weight.pos=0.1, loss\_weight.spec=1.0}
    \item \text{latent\_size=256, training.lr=0.01, loss\_weight.pos=0.1, loss\_weight.spec=10.0}
    \item \text{latent\_size=256, training.lr=0.01, loss\_weight.pos=1.0, loss\_weight.spec=0.1}
    \item \text{latent\_size=256, training.lr=0.01, loss\_weight.pos=1.0, loss\_weight.spec=1.0}
    \item \text{latent\_size=256, training.lr=0.01, loss\_weight.pos=1.0, loss\_weight.spec=10.0}
    \item \text{latent\_size=256, training.lr=0.01, loss\_weight.pos=10.0, loss\_weight.spec=0.1}
    \item \text{latent\_size=256, training.lr=0.01, loss\_weight.pos=10.0, loss\_weight.spec=1.0}
    \item \text{latent\_size=256, training.lr=0.01, loss\_weight.pos=10.0, loss\_weight.spec=10.0}
    \item \text{latent\_size=256, training.lr=0.001, loss\_weight.pos=0.1, loss\_weight.spec=0.1}
    \item \text{latent\_size=256, training.lr=0.001, loss\_weight.pos=0.1, loss\_weight.spec=1.0}
    \item \text{latent\_size=256, training.lr=0.001, loss\_weight.pos=0.1, loss\_weight.spec=10.0}
    \item \text{latent\_size=256, training.lr=0.001, loss\_weight.pos=1.0, loss\_weight.spec=0.1}
    \item \text{latent\_size=256, training.lr=0.001, loss\_weight.pos=1.0, loss\_weight.spec=1.0}
    \item \text{latent\_size=256, training.lr=0.001, loss\_weight.pos=1.0, loss\_weight.spec=10.0}
    \item \text{latent\_size=256, training.lr=0.001, loss\_weight.pos=10.0, loss\_weight.spec=0.1}
    \item \text{latent\_size=256, training.lr=0.001, loss\_weight.pos=10.0, loss\_weight.spec=1.0}
    \item \text{latent\_size=256, training.lr=0.001, loss\_weight.pos=10.0, loss\_weight.spec=10.0}
    \item \text{latent\_size=256, training.lr=0.0001, loss\_weight.pos=0.1, loss\_weight.spec=0.1}
    \item \text{latent\_size=256, training.lr=0.0001, loss\_weight.pos=0.1, loss\_weight.spec=1.0}
    \item \text{latent\_size=256, training.lr=0.0001, loss\_weight.pos=0.1, loss\_weight.spec=10.0}
    \item \text{latent\_size=256, training.lr=0.0001, loss\_weight.pos=1.0, loss\_weight.spec=0.1}
    \item \text{latent\_size=256, training.lr=0.0001, loss\_weight.pos=1.0, loss\_weight.spec=1.0}
    \item \text{latent\_size=256, training.lr=0.0001, loss\_weight.pos=1.0, loss\_weight.spec=10.0}
    \item \text{latent\_size=256, training.lr=0.0001, loss\_weight.pos=10.0, loss\_weight.spec=0.1}
    \item \text{latent\_size=256, training.lr=0.0001, loss\_weight.pos=10.0, loss\_weight.spec=1.0}
    \item \text{latent\_size=256, training.lr=0.0001, loss\_weight.pos=10.0, loss\_weight.spec=10.0}
    \item \text{latent\_size=384, training.lr=0.01, loss\_weight.pos=0.1, loss\_weight.spec=0.1}
    \item \text{latent\_size=384, training.lr=0.01, loss\_weight.pos=0.1, loss\_weight.spec=1.0}
    \item \text{latent\_size=384, training.lr=0.01, loss\_weight.pos=0.1, loss\_weight.spec=10.0}
    \item \text{latent\_size=384, training.lr=0.01, loss\_weight.pos=1.0, loss\_weight.spec=0.1}
    \item \text{latent\_size=384, training.lr=0.01, loss\_weight.pos=1.0, loss\_weight.spec=1.0}
    \item \text{latent\_size=384, training.lr=0.01, loss\_weight.pos=1.0, loss\_weight.spec=10.0}
    \item \text{latent\_size=384, training.lr=0.01, loss\_weight.pos=10.0, loss\_weight.spec=0.1}
    \item \text{latent\_size=384, training.lr=0.01, loss\_weight.pos=10.0, loss\_weight.spec=1.0}
    \item \text{latent\_size=384, training.lr=0.01, loss\_weight.pos=10.0, loss\_weight.spec=10.0}
    \item \text{latent\_size=384, training.lr=0.001, loss\_weight.pos=0.1, loss\_weight.spec=0.1}
    \item \text{latent\_size=384, training.lr=0.001, loss\_weight.pos=0.1, loss\_weight.spec=1.0}
    \item \text{latent\_size=384, training.lr=0.001, loss\_weight.pos=0.1, loss\_weight.spec=10.0}
    \item \text{latent\_size=384, training.lr=0.001, loss\_weight.pos=1.0, loss\_weight.spec=0.1}
    \item \text{latent\_size=384, training.lr=0.001, loss\_weight.pos=1.0, loss\_weight.spec=1.0}
    \item \text{latent\_size=384, training.lr=0.001, loss\_weight.pos=1.0, loss\_weight.spec=10.0}
    \item \text{latent\_size=384, training.lr=0.001, loss\_weight.pos=10.0, loss\_weight.spec=0.1}
    \item \text{latent\_size=384, training.lr=0.001, loss\_weight.pos=10.0, loss\_weight.spec=1.0}
    \item \text{latent\_size=384, training.lr=0.001, loss\_weight.pos=10.0, loss\_weight.spec=10.0}
    \item \text{latent\_size=384, training.lr=0.0001, loss\_weight.pos=0.1, loss\_weight.spec=0.1}
    \item \text{latent\_size=384, training.lr=0.0001, loss\_weight.pos=0.1, loss\_weight.spec=1.0}
    \item \text{latent\_size=384, training.lr=0.0001, loss\_weight.pos=0.1, loss\_weight.spec=10.0}
    \item \text{latent\_size=384, training.lr=0.0001, loss\_weight.pos=1.0, loss\_weight.spec=0.1}
    \item \text{latent\_size=384, training.lr=0.0001, loss\_weight.pos=1.0, loss\_weight.spec=1.0}
    \item \text{latent\_size=384, training.lr=0.0001, loss\_weight.pos=1.0, loss\_weight.spec=10.0}
    \item \text{latent\_size=384, training.lr=0.0001, loss\_weight.pos=10.0, loss\_weight.spec=0.1}
    \item \text{latent\_size=384, training.lr=0.0001, loss\_weight.pos=10.0, loss\_weight.spec=1.0}
    \item \text{latent\_size=384, training.lr=0.0001, loss\_weight.pos=10.0, loss\_weight.spec=10.0}
  \end{itemize}
\end{enumerate}

The default hyperparameters of ICSG3D were also adapted for the LCS6{\AA} dataset by searching for the best configuration from the list below.
\begin{itemize}
\scriptsize
  \item \text{batch\_size=10, latent\_size=1024}
  \item \text{batch\_size=10, latent\_size=128}
  \item \text{batch\_size=10, latent\_size=256}
  \item \text{batch\_size=10, latent\_size=512}
  \item \text{batch\_size=10, latent\_size=64}
  \item \text{batch\_size=1, latent\_size=1024}
  \item \text{batch\_size=1, latent\_size=128}
  \item \text{batch\_size=1, latent\_size=256}
  \item \text{batch\_size=1, latent\_size=512}
  \item \text{batch\_size=1, latent\_size=64}
  \item \text{batch\_size=20, latent\_size=1024}
  \item \text{batch\_size=20, latent\_size=128}
  \item \text{batch\_size=20, latent\_size=256}
  \item \text{batch\_size=20, latent\_size=512}
  \item \text{batch\_size=20, latent\_size=64}
  \item \text{batch\_size=40, latent\_size=1024}
  \item \text{batch\_size=40, latent\_size=128}
  \item \text{batch\_size=40, latent\_size=256}
  \item \text{batch\_size=40, latent\_size=512}
  \item \text{batch\_size=40, latent\_size=64}
  \item \text{batch\_size=5, latent\_size=1024}
  \item \text{batch\_size=5, latent\_size=128}
  \item \text{batch\_size=5, latent\_size=256}
  \item \text{batch\_size=5, latent\_size=512}
  \item \text{batch\_size=5, latent\_size=64}
\end{itemize}